# Citizen Science: An Information Quality Research Frontier

Roman Lukyanenko[1] · Andrea Wiggins[2] · Holly K. Rosser[2]

**Abstract**
The rapid proliferation of online content producing and sharing technologies resulted in an explosion of user-generated content (UGC), which now extends to scientific data. Citizen science, in which ordinary people contribute information for scientific research, epitomizes UGC. Citizen science projects are typically open to everyone, engage diverse audiences, and challenge ordinary people to produce data of highest quality to be usable in science. This also makes citizen science a very exciting area to study both traditional and innovative approaches to information quality management. With this paper we position citizen science as a leading information quality research frontier. We also show how citizen science opens a unique opportunity for the information systems community to contribute to a broad range of disciplines in natural and social sciences and humanities.

**Keywords** Citizen science · Information quality · Data quality · Crowdsourcing · Science · Humanities · Information systems

> "The urgency of environmental issues draws our attention to the management of finite resources, and the potential of digital tools to help us work with them effectively" (Light and Miskelly 2014, p. 14).

## 1 Introduction

Increased availability of digital resources to support critical decisions makes information quality (IQ) a paramount concern for organizations and society. Recognizing its social importance, information systems (IS) research considers IQ to be a core interest resulting in a long and rich research tradition (Batini et al. 2015; DeLone and McLean 1992; Madnick et al. 2009; Petter et al. 2013; Wahyudi et al. 2018; Wang and Strong 1996).

✉ Roman Lukyanenko
roman.lukyanenko@hec.ca

Andrea Wiggins
wiggins@unomaha.edu

Holly K. Rosser
hrosser@unomaha.edu

[1] Département de technologies de l'information, HEC Montréal, 3000, chemin de la Côte-Sainte-Catherine, Montréal, Québec, Canada

[2] University of Nebraska Omaha, 6001 Dodge St, Omaha, NE 68182, USA

While IQ remains a prominent subject in IS, much of foundational research on IQ has focused on traditional uses of data in corporate environments and for-profit organizations. In these settings information production was relatively transparent and controlled, typically created by known organizational actors (e.g., database operators) for well-defined purposes (Burton-Jones and Volkoff 2017; Wang 1998; Zuboff 1988). The growth of networks, including the Internet, fuels studies that investigate IQ in inter-organizational, trans-national projects and on the web (Alonso 2015; Batini and Scannapieca 2006; Calero et al. 2008; Heath and Bizer 2011; Wells et al. 2011; Zhu and Wu 2011). One such area of research deals with information created by ordinary people in the course of their daily lives, collectively known as user-generated content, or UGC (Doan et al. 2011; Levina and Arriaga 2014; Nov et al. 2014; Susarla et al. 2012). UGC encompasses such diverse and now ubiquitous forms as social media and social networks (e.g., Facebook, Twitter, Instagram, YouTube), online reviews (e.g., Yelp, TripAdvisor), blogs, chat rooms, discussion boards, comments (e.g., on news stories, products in an e-commerce catalog), knowledge production (e.g., Wikipedia, Wiktionary, Urban Dictionary), content sharing (e.g., Reddit, Digg), amateur news (e.g., Huffington Post), collaborative maps (e.g., OpenStreetMap), and others.

The growing societal importance of UGC is underscored by calls for its addition to economic calculations for national gross domestic product (Brynjolfsson and McAfee 2014). UGC gives organizations a unique view into the lives of ordinary people; more and more organizations consult UGC when designing new products, providing commercial services, shaping public policy, or conducting scientific research. Explicit open calls to members of the general public for specific information underlie the booming practice of crowdsourcing (Deng et al. 2016; Ghezzi et al. 2017; Love and Hirschheim 2017; Lukyanenko and Parsons 2018). To offer crowd services at massive scale, platforms like Amazon's Mechanical Turk and CrowdFlower.com allow companies to tap into thousands of paid crowd workers (Chittilappilly et al. 2016; Deng et al. 2016; Garcia-Molina et al. 2016; Li et al. 2016; Stewart et al. 2015). Organizations also create their own platforms, such as the widely popular Zooniverse (www.zooniverse.org), where scientists can post data processing tasks (e.g., image annotation, classification, transcription) to over a million engaged volunteers (Simpson et al. 2014). Data collection and processing through crowdsourcing is different from many other forms of UGC, as it focuses participation on specific, often standardized, information needs and may require specific background, interests, and abilities.

User-generated content stands to rapidly expand the scope of IQ research as it presents novel challenges and opens many new questions (Li et al. 2016; Lukyanenko and Parsons 2015b). Unlike information produced by organizational employees, suppliers, and contractors with close ties to the organization, UGC is created by "casual" content contributors with varying levels of subject matter expertise, reasons for producing content, backgrounds, and worldviews (Daugherty et al. 2008; Levina and Arriaga 2014; Lukyanenko et al. 2014b; Susarla et al. 2012). In some such projects, content producers remain largely anonymous, making it challenging to determine contribution quality based on traditional indicators like known credentials. Physical, and often cultural and emotional distance online further complicates the process of understanding and interpreting data (Deng et al. 2016; Gray et al. 2016). Lack of context of information creation, an important factor for appropriate use of information (Lee 2003; Shankaranarayanan and Blake 2017; Strong et al. 1997), can significantly complicate using UGC.

Many novel research questions emerge in response to the rise of UGC and crowdsourcing. For example, since UGC is typically produced voluntarily (as opposed to relying on extrinsic organizational rewards), new issues related to motivation emerge (Daugherty et al. 2008; Nov et al. 2014). Similarly, novel forms of information creation such as collaborative production (e.g., Wikipedia) or information co-creation, call for explanation and understanding of the mechanisms of quality control in these settings (Arazy et al. 2011, 2017; Liu and Ram 2011; Zwass 2010). Challenges involved in collecting, storing, and using UGC motivate innovations in database, data visualization, data integration and analytic solutions (Bonney et al. 2014a, 2014b; Daniel et al. 2018; Hochachka et al. 2012; Stevens et al. 2014).

Illustrating the potential of UGC to make significant contributions to the broader field of IQ, we focus here on one of its prominent types: *technology-supported citizen science*. Broadly, citizen science refers to participation of ordinary people in scientific research (Bonney et al. 2014b; Show 2015). The term describes a range of approaches to community-based science and public involvement in research (Newman et al. 2010, p. 1852). Although many labels exist to describe this phenomena and its more specific forms, citizen science has become the most popular (Kullenberg and Kasperowski 2016).

While citizen science has many interesting dimensions (e.g., scientific, social, educational, economic) (Burgess et al. 2017; Louv et al. 2012), we focus on its technological aspects. From the point of view of IS, citizen science involves an unusual type of technology when compared with traditional corporate IS, such as ERP or supply chain systems, but also differs to some extent from such technologies as social media (more on these differences below). In citizen science, data is created or processed by ordinary people, often (but not always) with no formal education in the subject matter to be used by scientists or domain experts who typically require information of highest quality.

A major aspect of online citizen science is the open and frequently anonymous participation, due primarily to large scales of participation, further complicating information quality activities. At the same time, as citizen science is increasingly part of modern scientific practice, it has attracted considerable attention from researchers from natural and social sciences, medicine, and engineering. For example, citizen science research, including aspects dealing with IQ is on the rise in ecology, biology, physics, chemistry, astronomy, geology, anthropology, and politics, as well as humanities (Fortson et al. 2011; Haklay 2013; Hand 2010; Show 2015; Stevens et al. 2014). Few other applications of IQ have received such intense interdisciplinary attention. Finally, although typically conducted in public sector organizations (NGOs, academia, government), aspects of citizen science have potential for translation to corporate research and development as well. Recent studies in information systems argue for the importance of conducting research in citizen science due to its implications beyond its immediate context (Levy and Germonprez 2017; Lukyanenko et al. 2016b).

Importantly, in dealing with multiple challenges of citizen science IQ, interdisciplinary researchers investigate novel theoretical concepts, and deploy and evaluate innovative IQ solutions that could potentially advance broader IQ thinking and practice. Motivated by this potential, we propose research on citizen science as a promising IQ research frontier. To better understand the potential of this domain, we review the IQ challenges in citizen science and relate them to traditional IQ concepts and approaches. We then offer directions for future research in citizen science and IS focusing on IQ and conclude by identifying the value that citizen science research can bring to strengthening connections between the information systems discipline and other fields in sciences and humanities.

## 2 Situating Citizen Science in the Broader UGC Landscape

Although citizen science is an established topic in natural sciences (below, we briefly discuss the history of citizen science), it is a relatively recent area of research in the information systems discipline, including a stream focused on information quality, and in related fields such as computing. Before discussing specific issues that relate to the quality of data in citizen science, we position citizen science in the context of other related phenomena. Table 1 provides definitions of some of related concepts (note: the list is not exhaustive, and is meant to show major categories to elaborate the niche which citizen science occupies from an IS perspective).

Citizen science largely relies on UGC, as volunteers are routinely engaged in collection, processing or analysis of data. However, citizen science brings an important scientific dimension that general UGC lacks. The scientific dimension heightens the importance of high information quality. Also, as citizen science projects tap into personal interest in public goods, successful projects can easily attract many engaged participants.

Many successful citizen science projects allow for high social interactivity among volunteers. As Silvertown (2010) noted in a *Nature* commentary, a solution to species identification by non-experts may lie in "social networking", whereby users are allowed to debate a particular case and among each other and vote up or down candidate species categories, precisely the mechanism now encoded in the increasingly popular iNaturalist platform (inaturalist.org) for biodiversity data contributions. Many scientific discoveries, such as that of Hanny's Voorwerp (see Table 2) on the Zooniverse platform, originated through the discussions on the forums provided to citizen scientists. However, high user interactivity may not be suitable for some citizen science projects (for reasons we discuss below), or can be simply unnecessary for the fulfilment of specific project objectives. Finally, the essence of social media is facilitation of social exchange. Issues of scientific discovery, high information quality, pursuit of objective truth, adherence to protocols and standards, and environmental conservation are very much on the fringes for most social media projects. As Maddah et al. (2018) argue, social media emphasizes personal experiences (or episodic information), rather than general or semantic information (the typical focus of citizen science).

Citizen science projects frequently "crowdsource" data collection or other activities. In this case, citizen science utilizes crowdsourcing as a work organization strategy. However, crowdsourcing first and foremost is a business model and as Levy and Germonprez (2017, p. 29) noted, "crowdsourcing is not rooted in citizens' intervening in the scientific process". Indeed, much of crowdsourcing has a prominent monetary dimension that is typically missing in citizen science. For example, turning to crowds for project funding or product development – crowdfunding – specifically pursues monetary goals (Choy and Schlagwein 2016; Zhou et al. 2018). Likewise, Amazon's Mechanical Turk and CrowdFlower.com serve as intermediaries between companies and online workers (Chittilappilly et al. 2016;

**Table 1** Examples of concepts closely related to citizen science

| Concept | Definition | Reference Sources | Scope of Phenomenon |
|---|---|---|---|
| User generated content | Information created online by the members of the general public, rather than organizational employees or other actors formally associated with an organization (e.g., suppliers) | (Daugherty et al. 2008, Levina and Arriaga 2014, Lukyanenko et al. 2014b, Susarla et al. 2012) | Diverse projects and platforms, such as social media, social networks, online reviews, blogs, chat rooms, discussion boards, comments, wikis, content sharing, amateur news, collaborative maps |
| Social media, social networks | An array of web-based technologies which allow direct interaction among users, including rapid communication and sharing of information | (Abbasi et al. 2018, Kane et al. 2014, Kapoor et al. 2018, Miranda et al. 2016) | Social networks, chats, message boards, forums, photo and video sharing apps and platforms |
| Crowdsourcing | A model of labour in which an organization turns to crowds (i.e., ill-defined members of the general public) to obtain desired data, goods, or services | (Brabham 2013, Ghezzi et al. 2017, McAfee and Brynjolfsson 2017, Poblet et al. 2017, Zhao and Zhu 2014) | Many business, public and even individual initiatives, such as crowdfunding, ideation, contest, citizen science, crisis management or online employment platforms |
| Collective intelligence | Knowledge that emerges as a result of collaboration between many individuals engaged in the same task | (Awal and Bharadwaj 2017, Bonabeau 2009, Malone et al. 2010, Woolley et al. 2010) | Problem-solving, decision-making, collaborative action in a variety of domains |

**Table 2** Examples of recent discoveries made by citizen scientists

| Discovery | Details of the discovery | Who facilitated the discovery?* | Source |
|---|---|---|---|
| Quasar ionization echo | Hanny's Voorwerp - an unknown cosmic body described as one of the most exciting recent discoveries in astronomy; discovered through the citizen science platform Zooniverse.org | Dutch school teacher, Hanny van Arkel | (Liu et al. 2014) |
| New pulsars | Over a dozen new pulsars, many with interesting properties, found using the platform Einstein@Home (einsteinathome.org) | Ordinary people across the world | (Clark et al. 2017, Liu et al. 2014) |
| Protein enzyme | Primate HIV enzyme folding solution discovered through the protein-folding project Foldit, advancing research on HIV treatment in humans | Ordinary gamers, such as "mimi" from UK | (Khatib et al. 2011) |
| Lost satellite | Located NASA's IMAGE satellite, "lost" in 2005 | Canadian amateur astronomer, Scott Tilley | (Voosen 2018) |
| Alien planets | Tens of exoplanets are discovered by citizen scientists in Planet Hunters (planethunters.org) | Ordinary people across the world | (Boyajian et al. 2016) |
| New beetles species | Several new species of beetles discovered in Borneo on a scientist-led citizen science trip | Eco-tourists | (Schilthuizen et al. 2017) |
| New spider species | New species of spider discovered by playing a mobile game, QuestaGame (questagame.com) that encourages observing the environment | Australian gamers | (Renault 2018) |
| New type of aurora | An auroral light structure different from the traditional auroral oval, called Strong Thermal Emission Velocity Enhancement (STEVE) to match its properties of subauroral ion drift; named by citizen scientist Chris Ratzlaff | Canadian amateur photographers, such as a club of "aurora chasers" from Alberta, Canada | (MacDonald et al. 2018) |
| Rediscovery of rare ladybug | Rare 9-spotted ladybug in a region where it was once common, but was believed to be extirpated | Elementary school student in New York State, US | (Losey et al. 2007) |
| Rediscovery of butterfly species | A colony of a rare butterfly, Stoffberg Widow, that had been believed extinct | Natural history society volunteers in Dindela, South Africa | (Lawrence 2015) |
| Tropical Atlantic fish species found in the Mediterranean | First documentation of the Sergeant major fish species in the Mediterranean, which is native to the Atlantic; such data is valuable for managing biological invasions | Recreational scuba diver along the coast of Tarragona, Spain | (Azzurro et al. 2013) |

*Note: Typically, citizens are the first to see the phenomena and alert the scientific community via the citizen science IS. However, scientists then commonly need to conduct rigorous testing or experimentation to confirm the discovery and publish it. Often, but not always, citizens are invited to co-author publications, such as the paper co-authored by Mardon Erbland, a citizen scientist, who discovered a new distribution of mosquito native to Asia in Eastern Canada (Fielden et al. 2015)

Deng et al. 2016; Garcia-Molina et al. 2016; Li et al. 2016; Stewart et al. 2015; Zhao and Zhu 2014). The commercial nature of many crowdsourcing initiatives creates a different information quality management environment compared to typical citizen science projects. In addition, many crowdsourcing projects are not interested in testing scientific theories or discovering ground truths, but rather engage online users in routine tasks that add business value. Crowdsourcing projects such as those on Amazon Mechanical Turk can frequently utilize redundancy or scale in the crowds to remove data outliers or obtain response distributions (Ipeirotis and Gabrilovich 2014; Tang et al. 2015; Zhao and Zhu 2014). This objective may be applicable to some crowdsourcing projects (e.g., mapping distribution of species in an area), but would not be aligned with the goals of making scientific discoveries, promoting environmental conservation, or raising awareness of scientific issues (Wiggins and Crowston 2011). Finally, crowdsourcing projects typically end when the desired data or services are obtained from the crowds; in contrast, citizen science projects are commonly long-term and involve continuous engagement with the contributors, including feedback, debriefing, and sharing results with interested participants (Newman et al. 2012; Rotman et al. 2012).

Citizen science also creates fertile grounds to leverage collective intelligence. Citizen science projects sometimes involve collaborative decision-making processes, creating opportunities to discover solutions or produce results superior to those which would have been created if volunteers or scientists worked individually (Malone et al. 2010). Citizen science can especially benefit from collective intelligence in projects that require group tasks, problem solving, and decision-making. For example, InnoCentive is an online platform which posts difficult research problems, such as how to synthesize certain chemical compounds, for crowds to solve. Best solutions can receive substantial monetary rewards, sometimes worth tens or hundreds of thousands of US dollars (Malone et al. 2010). However, citizen science activities such as environmental monitoring and conservation do not directly depend on collective intelligence for success.

As we show below, the differences between citizen science and other related fields translate into important nuances for information management and information quality improvement strategies. This makes it important to study information quality in citizen science in its own right. At the same time, the shared properties between citizen science and crowdsourcing, collective intelligence, and other forms of user generated content, mean that progress in information quality research on citizen science may also be leveraged to tackle challenges and problems related to information quality in related domains.

Finally, like citizen science, information quality research on crowdsourcing, user generated content, collective intelligence, social media and other emerging media forms, is a relatively recent subject in the information systems discipline (Kapoor et al. 2018; Lukyanenko and Parsons 2015b; Palacios et al. 2016; Tilly et al. 2016). Thus, focusing on citizen science – a societally important form of public-professional collaboration – can spearhead broader development of information quality research and continue to expand its scope and impact beyond the traditional research in the context of corporate organizational settings.

## 3 Why Study Information Quality in Citizen Science?

As we showed in the previous section, citizen science presents a unique study opportunity for information quality research community due to its distinctive forms of information production. In this section we focus on several important characteristics of citizen science that not only distinguish it from other types of UGC, but more importantly, make technology-supported citizen science particularly fertile for advancing IQ research.

### 3.1 Relevance to Science and Society

In recent decades, the IS research community has begun to explore relevant topics outside of traditional organizational settings (Levy and Germonprez 2017; Winter et al. 2014). Calls are increasing to conduct research in information systems that is explicitly concerned for large social groups and society (Goes 2014; Melville 2010; Seidel et al. 2013). Citizen science is uniquely positioned to establish IS as a discipline concerned with social progress and social improvement.

Studying IQ in citizen science stands to benefit sciences broadly and has potential for major social improvements (Burgess et al. 2017; McKinley et al. 2016). Much hope is vested in citizen science, which is most strongly adopted in sciences tackling humanity's "evil quintet" of climate change, overexploitation, invasive species, land use change, and pollution (Theobald et al. 2015). As Light and Miskelly (2014, p. 14) put it, "[t]he urgency of environmental issues draws our attention to the management of finite resources, and the potential of digital tools to help us work with them effectively." These are conversations that IS can ill afford to ignore in the Twenty-First Century.

Citizen science is also a booming hobby for online participants worldwide and a growing trend in science. In biodiversity research alone, it is estimated that more than two million people are engaged in major citizen science projects, contributing up to $2.5 billion of in-kind value annually (Theobald et al. 2015). As of 2009, the public was already providing 86% of biological survey data in Europe (Schmeller et al. 2009). Constantly pushing the boundaries of what is possible, scientists are engaging ordinary people in an ever increasing range of tasks. Volunteers may be asked to observe plants and animals (Hochachka et al. 2012), transcribe historical texts (Eveleigh et al. 2013), map cultural and geographic objects (Haklay and Weber 2008), identify natural resources to preserve and protect from development and exploitation (Vitos et al. 2013), design new drugs (Khatib et al. 2011), search for extraterrestrial intelligence (Korpela 2012), catalog galaxies (Fortson et al. 2011), describe oceans floors and document marine life (Pattengill-Semmens and Semmens 2003), track butterflies (Davis and Howard 2005), and report on active wildfires (Goodchild and Glennon 2010), among many other tasks. Citizen science can take volunteers into wilderness areas, as in the case of some biological monitoring projects, but are just as likely to occur in kitchens or backyards (some projects require mailing specimens, such as sourdough starters or soil samples), can be conducted entirely online (such as peering at photographs taken by the Hubble telescope to classify galaxies or scanned specimens to transcribe museum collections), or require little direct involvement, like in volunteer computing projects such as SETI@Home's search for extraterrestrial intelligence and a wide variety of other research projects supported by the BOINC distributed computing architecture (Hand 2010; Louv et al. 2012).

Ingenuous application of online technologies, coupled with enthusiasm and creativity of ordinary people, have already resulted in a number of high-profile scientific discoveries. Table 1 provides several examples of discoveries originating in citizen science projects.

Notably, citizen science information quality is already a well-explored subject in sciences (discussed further below). It builds on the long tradition of involving humans in data collection that invariably involved external research participants, field research settings, noisy and unreliable sources, and investments into innovative data collection infrastructure. Some date the origins of citizen science to Victorian-era Britain (or earlier) and online citizen science was in fact one of the early adopters of content producing technologies known as Web 2.0 (Bauer et al. 2000; Bonney et al. 2009; Brossard et al. 2005; Goodchild 2007; Louv et al. 2012; Osborn et al.

2005; Wiersma 2010). By bringing IS expertise to citizen science, IS researchers can contribute to advancing an established and socially valuable body of long-standing cumulative research. This work has potential to advance practice as a value multiplier.

### 3.2 Relentless Pursuit of Quality

Information quality is a core concern in science, and therefore also in citizen science. As Prestopnik and Crowston (2012b, p. 9) conclude from observing a real-world citizen science project development: "[scientists'] primary goal is that [citizen science projects] should produce large amounts of *very high quality data*. Virtually all other considerations are secondary" (emphasis added). Similarly, data quality is "[o]ne of the highest priorities" for the Atlas of Living Australia, which maps biodiversity of the continent based on citizen-supplied data, as well as many other sources, including government agencies (Belbin 2011). Of the challenges that face the practice of citizen science, information quality is consistently considered paramount (Cohn 2008; Kosmala et al. 2016; Lewandowski and Specht 2015; Sheppard and Terveen 2011; Wiggins et al. 2011).

The persistent and often lopsided focus on IQ in citizen science contrasts with many other types of UGC (e.g., social networking, micro-task markets, blogs), where information quality competes with other priorities for attention, such as profit, enjoyment of participants, ability to advertise products, disseminating content, attracting followers, co-creating products, and solving practical problems (Ipeirotis et al. 2010; Levina and Arriaga 2014; Sorokin and Forsyth 2008; Susarla et al. 2012; Wattal et al. 2010; Zwass 2010). In contrast, in the context of science, these considerations are present but typically draw less attention because they are less core to the success of scientific research and are generally considered to be operational issues. Empirical evidence lies at the heart of the scientific method, making information quality a critical and central citizen science theme. Yet, this persistent interest in IQ is what makes citizen science an exciting setting for better understanding and advancing the general information quality research tradition.

As with any complex social phenomena, citizen science is more than just data collection and information quality. It also includes considerations of the engagement of volunteers in doing research; increasing public trust in the scientific enterprise; co-creation of scientific endeavors; engagement of the public in policies and debates arising as a result of science; building communities; involving people in addressing climate change and environmental sustainability; fostering education, literacy, and awareness of science among the general population (Bonney et al. 2009; Irwin 1995; Irwin and Michael 2003; Levy and Germonprez 2017; Sieber 2006). While these are important to the phenomenon and practice of citizen science, conducting research in information quality in citizen science holds undisputed value among scientists, and IS researchers are particularly well positioned to conduct research affecting the core issue in this emerging field.

### 3.3 A Multitude of Wicked Challenges

IS researchers are frequently challenged to embark on risky research that tackles difficult and ill-defined, wicked problems (Hevner et al. 2004). Due to its unique nature, citizen science has presented a number of information quality research problems that are either unreported in traditional organizational settings or familiar issues with notable contextual considerations.

One of the interesting aspects of citizen science is its typically democratic and open nature (Irwin 1995). Citizen science is fundamentally about empowering ordinary people in advancing research endeavors. Consequently, it is against the spirit of citizen science to discriminate against participants based on lack of scientific training or low literacy (ESCA 2015). Citizen science also puts an important premium on unique local knowledge. Yet it is precisely these characteristics that may endanger information quality (Gura 2013):

> "You don't necessarily know who is on the other end of a data point … It could be a retired botany professor reporting on wildflowers or a pure amateur with an untrained eye. As a result, it is difficult to guarantee the quality of the data."

In many emerging crowdsourcing environments, such as micro-task markets (Garcia-Molina et al. 2016; Lukyanenko and Parsons 2018; Ogunseye et al. 2017; Palacios et al. 2016), one of the ways to deal with information quality is to restrict participation to those members of the community deemed competent to produce high-quality information (Allahbakhsh et al. 2013; Chittilappilly et al. 2016; Ogunseye et al. 2017). While this is technically also possible to do in citizen science, it both runs against the spirit of openness and inclusion and can jeopardize the recruitment and retention of contributors that is key to success in projects that rely on voluntary labor. Yet, as noted earlier, high-quality consistent data is needed to make strong inferences and conclusions. Thus, scientists face a dilemma of optimizing information quality while keeping participation levels high enough to meet research needs and remaining sensitive to all relevant stakeholders. The challenge therefore becomes to preserve the openness of citizen science, allowing anybody interested to participate while delivering information of sufficient quality for scientific research.

Another consequence of open and often anonymous participation is that citizens often have diverse backgrounds and hold different views from each other. Sometimes the differences between scientists and contributors as well as the nature of the information are truly "extreme," as in the case of a conservation project in the Congo rainforest basin where data collectors include non-literate seminomadic hunter-gatherers (Pygmies), local farmers, loggers, and foresters (Rowland 2012; Stevens et al. 2014; Vitos et al. 2013). Lukyanenko et al. (2014a, 2014b) show that when views of volunteer contributors differ from those of professional experts (scientists) and one another, quality of contributions may suffer.

For the reasons above, quality remains a wicked problem. Indeed, concerns about IQ are so ingrained that scientists continue to face challenges in sharing findings based on citizen data with the scientific community even when they can conclusively demonstrate that data quality is high (Law et al. 2017; Lewandowski and Specht 2015; Riesch and Potter 2013). It is this untamed nature that makes citizen science information quality research interesting and productive.

## 4 Where Is IS in the Citizen Science Landscape?

Despite the growing societal value of citizen science and many opportunities for IS researchers, IS continues to lag behind such disciplines as biology and education in working with citizen science as a context for research. To assess the extent of the gap and also estimate the general volume of citizen science research, we conducted a review of citizen science literature and present the results below.

Figure 1 shows the growth of research on and with citizen science across disciplines and time. To generate this analysis, we used exact-match querying for the phrase "citizen science" occurring in the title, abstract, or keywords of journal articles and conference proceedings indexed in Scopus (3002), Web of Science (2302), IEEE eXplore (130), and ACM Digital Library (204). This is a conservative method for identifying the extant literature on citizen science, as not all relevant publications use this currently popular terminology, but our initial query results included 5806 items. The results were manually de-duplicated by using title name and DOI for reference, since there was extensive cross-indexing in several sources. We also removed editorial introductions to journal issues, dissertations, workshop papers, and posters to leave only full papers. Abstracts were then examined for papers with titles that were not obviously relevant to citizen science to evaluate the extent of focus on citizen science as a phenomenon or method. Papers mentioning citizen science only in passing, e.g., as an example of a broader trend (as opposed to a primary focus of the work) were eliminated, as were a few papers for which DOIs did not resolve during verification, bringing the total corpus to 1633.

The resulting papers were categorized to discipline and sub-discipline based on publication venue using the Classification of Instructional Programs (CIP) taxonomy by the United States Department of Education, National Center for Education Statistics.[1] Although any such classification scheme is imperfect and the CIP is pending revision in 2020, nearly all sub-specialties were located within the document and could be associated with a relevant discipline. Since some publication venues are non-disciplinary, e.g., *Public Library of Science*, papers published in these journals were categorized to discipline based on title and keywords, where available, or by checking the abstract if necessary. To avoid underestimating the volume of IS research and be more conservative, all papers published in the Hawai'i International Conference on Systems Science (HICSS) were classified as "IS" although work originating in Computing (Computer Science) and other disciplines is also often presented in this venue. For the sake of effective summary, several fields with just one or two publications were folded into "nearest neighbor" disciplines; these included Mathematics, Leisure studies, Linguistics, Philosophy, Psychology, and Business. Sub-disciplines (or primary themes, where disciplinary specialties were not present) were assigned to 1545 papers (95%) as several of the disciplinary categories are very broad, e.g., biology.

As Fig. 1 clearly shows, there has been substantial growth in annual publications on the topic of citizen science over the last 30 years. Even in our highly conservative data set, for the fields with more than two publications (over all time), IS leads only Agriculture, Architecture (i.e., Urban Planning), Humanities, and Legal scholarship. The closely related fields of Biology and Natural Resources Conservation together yielded 819 papers, approximately 50% of the total, and Ecology was the leading sub-discipline with 257 papers. By comparison, Computing had 260 (primarily empirical) papers while IS had just 17, and notably, the papers in IS originated with a very small handful of authors and several are non-empirical in focus.

It is further notable that the range of disciplinary specialties represented within Computing was not entirely out of line with the interests for IS scholars; Fig. 2 shows the sub-disciplinary topic areas for Computing publications, drawn primarily from the specific publication venues, e.g., ACM conferences on pervasive and ubiquitous computing (Ubicomp) or human-computer interaction (Design and Users). Of these, topics of interest for IS included Information Quality, Users, Ubiquitous Computing, Crowdsourcing, Social Computing, and Design. Among IS publications, the primary themes included Design (7), Information Quality (3), Social Computing (2), Crowdsourcing (1), Games (1), Open Source Software (1),

[1] https://nces.ed.gov

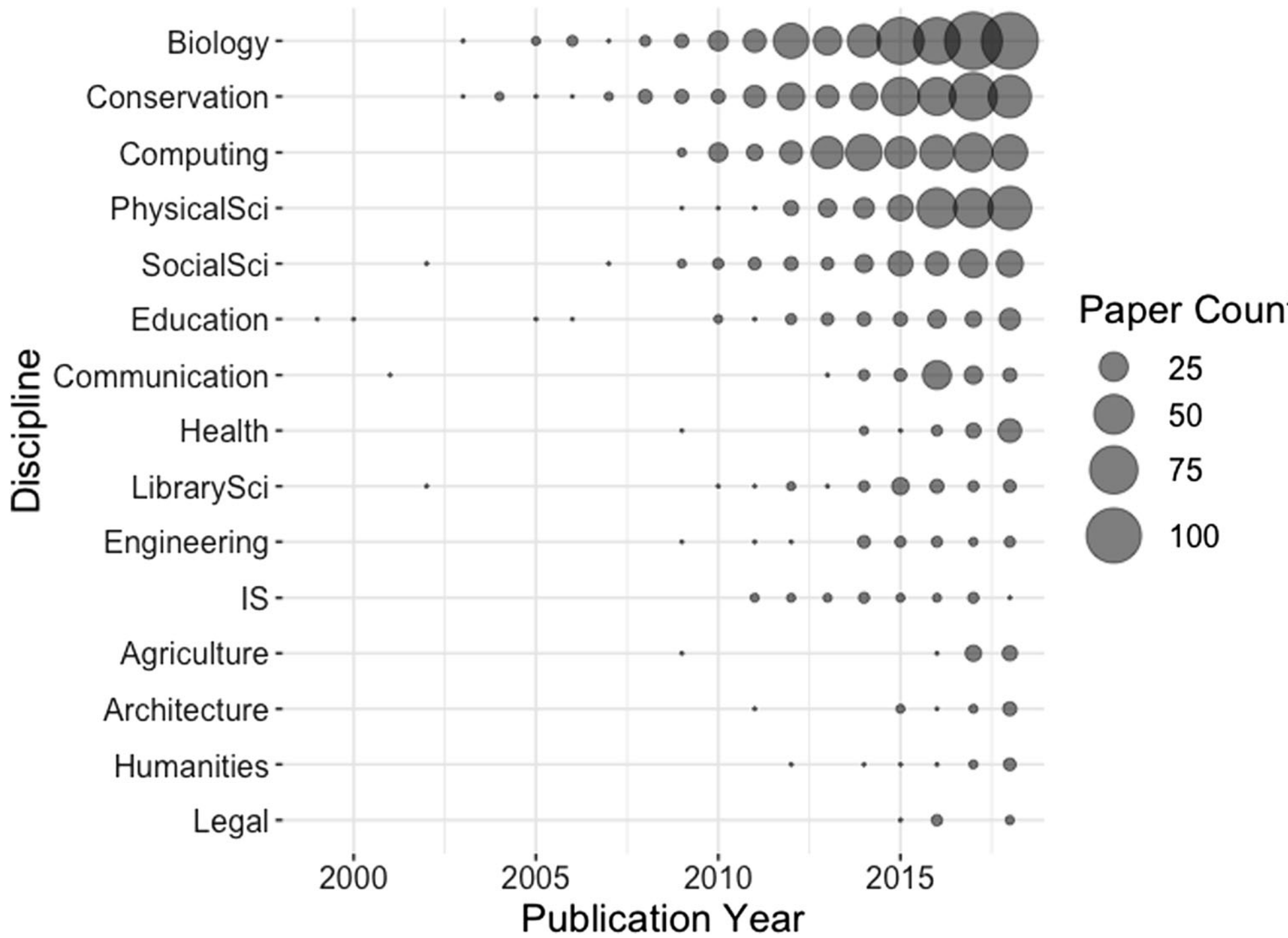

**Fig. 1** Publications on citizen science by discipline over time

Users (1), and a general perspective article (Levy and Germonprez 2017). Across the entire data set, the topic of Information Quality appeared 49 times, including 29 Conservation papers where it was a more common primary focus than species groups or habitats, and 15 papers in Computing, although Information Quality was in fact a dominant theme in far more publications than was captured by this analysis.

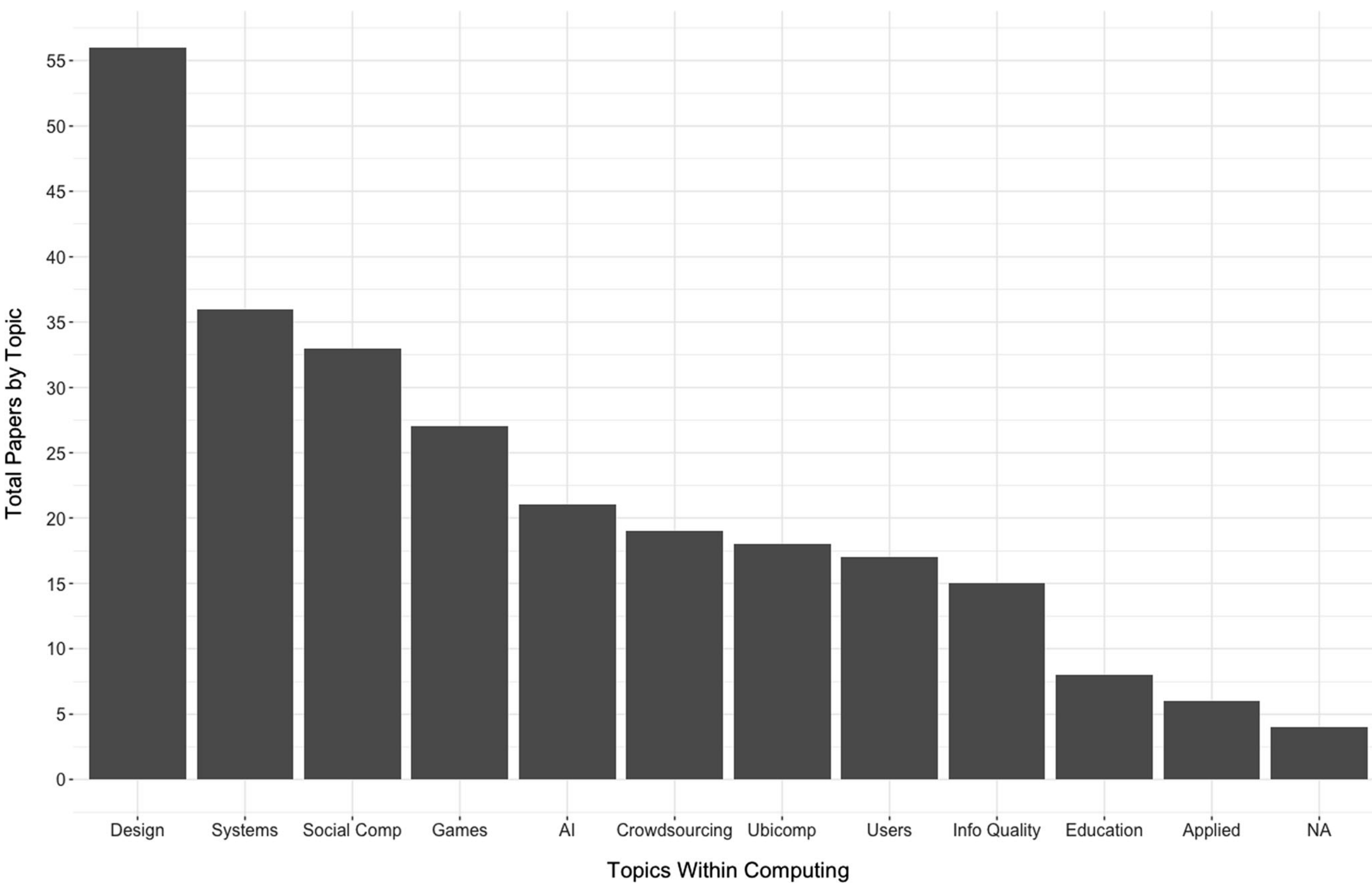

**Fig. 2** Sub-disciplinary areas of focus for papers on citizen science in Computing (NA designating opinion and perspective work)

As we can see, developing and applying information systems theories to improve praxis of citizen science offers an as yet unexplored opportunity for IS researchers to have impact beyond its traditional boundaries. Through citizen science, IS researchers can begin to address an important call to expand the frontiers of information systems research beyond traditional core subject matter and establish itself in the broader constellation of sciences (Beath et al. 2013) as well as doing a major service for the public good.

## 5 Information Quality Research Opportunities in Citizen Science

While information quality in citizen science is already an active area of research in sciences, it has only recently captured the attention of IS researchers (Lukyanenko et al. 2014b; Nov et al. 2014; Robinson and Imran 2015; Sheppard et al. 2014; Wiggins and Crowston 2011; Wiggins and He 2016). Interdisciplinary researchers focused on both traditional approaches to information quality, as well as implementing novel, untried methods and techniques have studied citizen science. The findings generated in this context can offer interesting and original insights with promise to advance the general understanding of theoretical and practical nature of IQ. Below, we illustrate this potential by examining major trends in citizen science information quality research.

### 5.1 Training and Evaluating Learning and Performance

Information quality research in organizational contexts studies many factors to improve IQ, including total data production monitoring and control, adherence to best practices in IS development (e.g., requirements elicitation, participative design, conceptual, database and interface design), training of data creators (e.g., database clerks), clarity of data entry instructions, close monitoring of the data production process, providing timely quality feedback and continuous improvement of information production processes (Ballou et al. 2003; Batini and Scannapieca 2006; Lee et al. 2002; Levitin and Redman 1995; Redman 1996; Strong et al. 1997; Wang et al. 1995a, 1995b). Many of these strategies have been applied and evaluated in citizen science projects (Bonney et al. 2009; Kosmala et al. 2016; Lewandowski and Specht 2015; Newman et al. 2003; Wiggins et al. 2011). Surveys of citizen science projects show that multiple quality assurance and quality control strategies are usually implemented (Wiggins and Crowston 2014). Thus, citizen science is already a partial extension of the IQ tradition in traditional settings.

Much like data entry in organizations, some citizen science projects require testing or certification in a certain skill to participate (Crall et al. 2011; Lewandowski and Specht 2015). Often projects naturally attract or are specifically marketed to members of the general public with domain expertise, such as avid birders (e.g., Cohn 2008). Insisting on high domain expertise or prequalifying tests of expertise can severely curtail participation in many projects and runs contrary to the open spirit of citizen science (Lukyanenko et al. 2014b; Wiggins et al. 2011). Alternative approaches, such as embedded assessments that evaluate learning or skill through the process of participation as opposed to using a quiz or other explicit instrument, are currently a hot topic of research in evaluation related to citizen science (Becker-Klein et al. 2016; Nelson et al. 2017; Ogunseye and Parsons 2016). A typical embedded assessment task for an online content classification project injects a handful of images into the workflow that have been pre-classified by experts to evaluate contributors' agreement with experts, producing individual-level performance measures.

Another traditional measure to improve IQ is training for data creators. Training has been widely applied and studied in the citizen science context. Anecdotal and direct evidence of the positive impact of training on accuracy is reported (Delaney et al. 2008; Newman et al. 2003; Wiggins et al. 2011). Citizen science studies explore theoretical boundaries on how training is designed and delivered, identifying limitations in projects with many users, globally distributed participants, ill-defined tasks, minimal resources for training, or large and complex domains (e.g., entirety of natural history). Furthermore, requiring extensive training may discourage participation, can lead to hypothesis guessing, and may result in other biases (Lukyanenko et al. 2014b; Ogunseye et al. 2017). For example, participants who know the objective of the project may overinflate or exaggerate information (Galloway et al. 2006; Miller et al. 2012). Some have explored the use of tasks which specifically do not require training (Eveleigh et al. 2014; Lukyanenko et al. 2019), while others have investigated the possibility of attracting diverse crowds so that training induced biases are mitigated due to the diversity of the participants (Ogunseye et al. 2017; Ogunseye and Parsons 2016). All these are novel ideas for traditional information quality research. As training is prevalent in corporate settings, findings from citizen science may suggest new approaches for this context as well.

### 5.2 Optimizing Tradeoffs

In citizen science, as in broader IQ research, information quality is considered a multidimensional construct composed of such dimensions as accuracy, timeliness, completeness (Arazy and Kopak 2011; Ballou and Pazer 1995; Girres and Touya 2010; Haklay and Weber 2008; Lee et al. 2002; Nov et al.

2014; Rana et al. 2015; Sicari et al. 2016). One of the findings of traditional IQ research are the trade-offs between dimensions (Batini and Scannapieca 2006; Pipino et al. 2002; Scannapieco et al. 2005). For example, previous research in corporate settings proposed trade-offs between consistency and completeness (Ballou and Pazer 2003), and accuracy and timeliness (Ballou and Pazer 1995). These findings have been corroborated in the citizen science context as well.

For example, research established a common trade-off in citizen science between accuracy and data completeness (Parsons et al. 2011). Specifically, to maximize accuracy, projects may be tempted to create restrictive data collection protocols and limit participation to the most knowledgeable contributors. While this strategy leads to increased accuracy of classification (e.g., identification of species), it also limits the amount of data (or completeness) such projects may otherwise achieve. A pressing challenge in citizen science is finding ways to mitigate such trade-offs, allowing many people to participate and provide information in an unconstrained manner, while ensuring that the information provided is adequate for scientists to use in research.

A novel trade-off between consistency and discovery appears to be further inherent to the citizen science context. Consistency is an important dimension of citizen science quality as it promotes reliability and comparability of results, often requisite for scientific data analysis. For example, in the River Watch Project (www.iwinst.org), "it is critical that the data is collected in exactly 'the same way, every time' so that any differences that are found are real and not due to inconsistent procedures. The goal is to minimize 'variables' as much as possible, so that data can be meaningfully compared." (Sheppard and Terveen 2011, p. 33). Researchers in citizen science have applied analysis techniques such as confusion matrices, finding that contributors who are consistently wrong can be just as valuable as those who are consistently right for certain tasks, because consistent responses can be corrected to make the data useful (Simpson et al. 2013).

Such standardization, of course, may preclude some projects from attaining another important scientific objective, discovery, where variability is a potential requisite (Lukyanenko, Parsons and Wiersma 2016, Nielsen 2011). Recent work in IS began to examine whether using innovative instance-based approaches to data collection and storage can eliminate trade-offs that appeared to be inevitable in citizen science (Parsons et al. 2011). An instance-based approach to storage and collection focuses on capturing unique attributes of individual objects of interest (e.g., birds, plants, geological and geographic forms, events), rather than their similarity. This information that is inherently variable and sparse can then be made consistent through the application of high performance data mining algorithms and inferential rules. Recent experiments demonstrate the advantages of this approach over traditional methods of data collection and storage in increasing simultaneously accuracy, completeness and discovery (Lukyanenko et al. 2014b, a), however the ability of this approach to promote consistency is unknown, suggesting a need for additional research in this direction (Lukyanenko, Parsons, Wiersma, et al. 2016). Other analytical techniques have also been applied, such as using effort information to standardize data (Bird et al. 2014), or using modeling (Sousa et al. 2018; Strien et al. 2013), applying different analytic approaches to solving some of the problems in participant variability, which also may have utility for IS research. Considering this work, there is a clear need for further research on dimensions and the nature of IQ tradeoffs in citizen science.

As illustrated by the examples above, a good deal of citizen science research is quite similar to the typical work information systems scholars have conducted in other contexts. This makes the findings from citizen science potentially transferable to broader areas of IQ and allows it to contribute by corroborating the existing theories and implementing proposed approaches. Yet another exciting aspect of citizen science is a growing number of ideas and strategies that are original and unique, promising novel insights and opportunities for the broader information quality research tradition. We focus the remainder of this section on select contributions that advance or challenge established theoretical notions and offer innovative practical solutions.

### 5.3 Human Factors in IQ

An important contribution of citizen science context is better understanding of individual human factors' impact on IQ. In traditional organizational settings, common training, background, instructions, and organizational culture and norms resulted in a relatively homogenous views of people in the same roles. Similarly, data creators in corporate environments are compelled to create high quality data as part of the general disposition to perform work well, despite known individual differences, e.g., in virtual teams (Watson-Manheim et al. 2002). Consequently, research on individual factors in IQ has been scarce. In a recent "information quality research challenge paper", Alonso (2015) laments: "[m]ost of the data quality frameworks … superficially [touch] the quality of work that is produced by workers. New labeling tasks that rely heavily on crowdsourcing need to pay attention to the different social aspects that can influence the overall quality of such labels. Payments, incentives, reputation, hiring, training, and work evaluation are some of those aspects" (p. 2).

By contrast to employees in corporate settings, the voluntary nature of content creation in citizen science means we must assume variation in backgrounds, motivation, knowledge, literacy skills, and other abilities, which motivates increased attention to the impact of individual factors on IQ. Many studies have demonstrated the importance and positive impact of motivation on quantity of contributions, noting the particular importance of intrinsic motivation in citizen science (Flanagin and Metzger 2008; Nov et al. 2011; Prestopnik et al. 2013). Nov et al. (2014) report positive impact of collective motives (contribution to collective goals) and rewards, such as increased reputation among project members and beyond or making new friends, on accuracy of contributions.

The nature of citizen science data collection also introduces biases in data that are generally assumed not to be present when information production occurs in controlled settings. These include self-selection biases, predisposition of volunteers to notice charismatic, unusual, easy to report events and objects, and spatial and temporal biases (Janssens and Kraft 2012), all of which are also known problems in professional science (Kosmala et al. 2016). Citizen science researchers have documented, investigated, and proposed solutions to detect and mitigate these biases when analyzing data and drawing conclusions (Crall et al. 2011; Delaney et al. 2008; Girres and Touya 2010; Wiersma 2010). These ideas could significantly enrich IS quality research, where individual variation in information production has attracted attention only recently (Burton-Jones and Volkoff 2017), and assumptions of bias, or lack thereof, may merit re-examination.

### 5.4 Hardware Factors

Data collection in traditional organizational settings was typically conducted in a well-controlled environment. Among other things it meant that in designing data collection process, IS specialists decided the hardware available in the process of design. In citizen science, controlling which devices volunteers use is much more problematic (He and Wiggins 2017). Citizen science research stands to enrich information quality literature by examining the impact of hardware and devices as well as the role of physical settings for doing data entry (Eveleigh et al. 2014; Stevens et al. 2014; Wiggins and He 2016), factors that have not received much attention in traditional IQ research. Among other things, research in citizen science has underscored the importance of designing for multiple devices and creating processes which are simple, flexible, easy to use, and work on devices with memory and Internet connectivity constraints (e.g., mobile apps or miniaturized systems) (Cohn 2008; Higgins et al. 2016; Kim et al. 2013; Newman et al. 2010; Prestopnik and Crowston 2012a). Another interesting idea, rarely investigated in corporate settings, is to allow users to explore, experiment, tinker, and even “fail” when working with the systems as a strategy for identifying ways to develop more robust and usable tools (Newman et al. 2010).

A general lesson emerging from this research is that the physical environment and hardware characteristics have a measurable impact on information quality – a novel consideration for traditional information systems research, which routinely abstracts away implementation details when dealing with representational issues (Jabbari et al. 2018; Jabbari Sabegh et al. 2017; Wand and Weber 1995). This insight may be increasingly relevant in “bring your own device” and app-driven organizational settings (French et al. 2014; Hopkins et al. 2017; Jordan 2017; Weeger et al. 2015).

### 5.5 Social Factors

Another original contribution of citizen science is the exploration of social factors in content creation. Social networking and interactivity between participants is a common design feature of citizen science IS (Haklay and Weber 2008; Silvertown 2010). Yet, unlike general social networking projects (such as Facebook, Twitter, Youtube), in citizen science these features are implemented and investigated with explicit information quality objectives, making this a particularly interesting context for understanding the implications of sociality on information production (Nov et al. 2014; Silvertown 2010). Silvertown (2010) suggested that social networking could aid novices in species identification through the process of peer identification, which Wiggins and He (2016) examined in a study of community-based data validation, where social practices and device types used to collect information impacted information quality. Social networking is generally expected to improve information accuracy, foster peer-based learning and co-creation where participants are able to both generate and evaluate content, develop new knowledge, and refine existing skills, all of which can lead to higher quality UGC (Lukyanenko 2014; Silvertown 2010; Tan et al. 2010). Contributors remain aware of the social context in which observations are being made: the study by Nov et al. (2011) found the impact of social motives on quantity of observations.

As the same time, overreliance on social media in the context of citizen science may impose scientific limitations and negative impacts on quality. High user interactivity may lead to group think or hypothesis guessing. As scientists frequently conduct online experiments in the context of citizen science (e.g., Lukyanenko et al. 2014a), high social interactivity may engender treatment spillover, when participants in one condition may communicate with the participants in another (e.g., treatment or control) condition. Social media also brings additional challenges of managing a less structured user engagement, which can be fraught with cyberbullying, trolling or other malignant user behavior prevalent on social media platforms (Kapoor et al. 2018), and may undermine citizen science projects.

As social networking has only recently been incorporated into corporate information production (Andriole 2010), there is scarce work on IQ implications of these platforms, and findings from citizen science may offer useful insights to social networking-related IQ research in corporate settings.

Consider malfeasance, a persistent concern in open online settings (Law et al. 2017), where employing social networking has shown promise for data sabotage detection (Delort et al. 2011; Wiggins and He 2016). Research has also considered features of systems that make them more robust to malfeasance (Wiggins and He 2016), as anecdotal evidence has suggested extremely low levels of malicious behavior in citizen science. These lessons may be applicable in more traditional settings to detect data sabotage by disgruntled employees. Research on social factors in citizen science can also contribute to the emerging area of online digital experimentation conducted for marketing or product development purposes in commercial contexts (e.g., Crook et al. 2009; Davenport and Harris 2017; Kohavi et al. 2013).

### 5.6 Re-Examining IS Research Assumptions

Citizen science sheds new light on and invites reconsideration of several long-standing assumptions of IS research. Some researchers argue that novel challenges of understanding and improving quality in citizen science warrant extending prevailing definitions of IQ and suggest rethinking traditional methods of improvement (Lukyanenko and Parsons 2015b). The organizational IQ paradigm focused on the data consumers and has defined quality as fitness of data to their needs (Madnick et al. 2009; Wang and Strong 1996). This view remains central to much of citizen science, as scientists routinely approach citizen science the same way they approach scientific research: they formulate predefined hypotheses, set goals for the projects, and then use the data provided by the volunteers to corroborate their predictions and assumptions (Lukyanenko and Parsons 2018; Wiersma 2010). This strategy is also sound as it reduces noise and variation in the data and makes the resulting data much easier to interpret and analyze.

Notwithstanding the value of approaching citizen science from the traditional fitness for use perspective, it may have limitations for some projects. A new definition of IQ has been proposed defining it as "the extent to which stored information represents the phenomena of interest to data consumers and project sponsors, as perceived by information contributors" (Lukyanenko et al. 2014b, p. 671). This definition avoids the traditional "fitness for use" definition (Madnick et al. 2009) since contributors may struggle to provide data that satisfies data consumer needs and may not even be fully aware of them. Thus, holding data contributors to data consumer standards may curtail their ability to provide high quality content as defined by data consumers, or suggest a need to refine instructions, procedures, or expectations. Guided by this definition, a series of laboratory and fields experiments demonstrated that accuracy and completeness of citizen science data can indeed be improved by relaxing the requirements to comply with data consumer needs (Lukyanenko et al. 2014a, 2014b; Lukyanenko et al. 2019), which is a standard project design strategy for achieving data quality targets in citizen science. Further, for projects focusing on new, emerging phenomena, it may be challenging to anticipate the optimal structure of citizen science data due to the different needs of diverse data consumers, so traditional solutions for storage premised on a priori structures (e.g., relational databases) may be inadequate in this setting (Sheppard et al. 2014). The new definition, however, remains controversial: the original definition was clearly useful for project leaders, and the data resulting from putting the views of data contributors first may be sparse and difficult to work with or even intractable for scientific requirements. Nonetheless, the advantages for some contexts suggest that the traditional fitness for use view may not fully capture the diverse and changing IQ landscape. These findings further provide a novel connection between conceptual modeling grammars (i.e., rules for building conceptual models and constructs that can be used to do so, see Gemino and Wand 2004, Mylopoulos 1998, Wand and Weber 2002, Woo 2011), database design approaches, and IQ, which are traditionally considered quite different research domains. The citizen science context revealed an exciting new possibility for combining research strands within IS.

Another controversial idea born out of citizen science is the notion that expertise may be harmful to IQ (Ogunseye and Parsons 2016). Contrary to the prevailing common sense assumption, Ogunseye and Parsons argue that it is the essence of expertise to be focused, and potentially ignore what is deemed irrelevant based on prior knowledge. Citizen science is a setting where discoveries are possible, and thus, it is important to have an open mind (Lukyanenko et al. 2016a). Crowdsourcing strategies where added value is premised on diverse contributions, e.g., open innovation (Brabham 2013; Howe 2008), are a prime example of this principle.

An active stream of research has studied design antecedents of quality in citizen science: many novel IS design solutions are rapidly tested with information quality being the most common outcome variable evaluated. Due to the importance of participants' motivation in creating large quantities of accurate information, design antecedents focused on motivation have received the most attention, including design features that support promotion of collective goals, ability to interact and make friends, rewards, and game-like features (Elevant 2013; Eveleigh et al. 2013; Nov et al. 2011; Prestopnik et al. 2013; Prestopnik and Crowston 2011). It can also offer a setting for research on participatory design (Bowser et al. 2017; Lukyanenko et al. 2016b; Preece et al. 2016), another major IS topic (Bodker 1996; Bratteteig and Wagner 2014; Garrity 2001; Klein and Hirschheim 2001; Kyng 1995).

Growing evidence demonstrates benefits of gamification for data quantity and discovery. For example, Foldit is a highly successful project that turns a complex and esoteric task of designing proteins into an interactive game resulting in important discoveries in healthcare and biology (Belden et al. 2015; Khatib et al. 2011; Khoury et al. 2014). Defying the notoriously mysterious nature of quantum theory, the Quantum Game Jam project (scienceathome.org) effectively engages ordinary people in such tasks as solving the Schrödinger equation. As gamification is quite novel in IS (e.g., Amir and Ralph 2014; Richards et al. 2014) citizen science can pave way for wider adoption of gamification in other settings. At the same time, gamification has potential to subvert IQ by incentivizing "gaming the system" and irritating or demotivating some of the most prolific and reliable contributors (Bowser et al. 2014), suggesting both that the benefits of gamification are audience specific, and that additional work is needed on this topic.

The quality challenges in citizen science also fuel the search for innovative design solutions, such as applying artificial intelligence techniques like machine learning. Artificial intelligence appears fruitful where knowledge and abilities of volunteers are deemed limited (Bonney et al. 2014a, 2014b; Hochachka et al. 2012). Typically, however, machine learning is applied post hoc – after data is captured – to detect trends or label items based on data provided (Hochachka et al. 2012; Tremblay et al. 2010). Citizen science is one of the pioneers of artificial intelligence used to augment and improve quality at the point of data creation. For example, the Merlin App aids people with bird species identification based on uploaded photographs and other basic information, using data from eBird contributors to predict likely species matches. The decision is made in real time and allows people to confirm or reject the automatic recommendation; the machine learning results can also be used to dynamically generate more effective data collection forms. The same project uses post hoc data mining to sift through millions of bird sightings to identify outliers, automatically flag erroneous species identifications, and detect potential vandalism. The substantial progress in the application of artificial intelligence and high-performance computing as well as growing evidence of effective adaptation of successful project designs led to recent *Nature* (Show 2015) and *Science* (Bonney et al. 2014a, 2014b) articles making the cautious conclusion that the problem of data quality in citizen science is "becoming relatively simple to address" (Show 2015, p. 265) for some projects. Such a conclusion, of course, is based on limited indicators of quality based on accuracy of identification (e.g., of biological species) which ignores many other potentially relevant dimensions (Lukyanenko et al. 2016a). At the same time, such signals from the leaders in the citizen science community are quite telling of the significant progress the field made in a very short period of time, which also stands ready to advance the broader field of IS.

Finally, a notable contribution of the citizen science context is investigating IQ in real settings. As visitors of websites can be easily converted into research subjects by assigning them to different control and treatment conditions (Lukyanenko and Parsons 2014), many citizen science experiments can be conducted in field settings (Delaney et al. 2008; Lewandowski and Specht 2015). Despite challenges in controlling for confounds, field experiments offer several advantages over laboratory experiments or surveys. Using a real setting allows tracking real user behavior (as opposed to behavioral intentions). It also allows for research with the population of interest rather than surrogates (e.g., students). Conducting research in a real setting also increases external validity compared to similar studies conducted in laboratory environments. As field experimentation is generally scarce in IS (Siau and Rossi 2011; Zhang and Li 2005), citizen science field experimental evidence can be used to corroborate and compare findings from more prevalent surveys and laboratory data (e.g., Jackson et al. 2016).

### 5.7 Summary

Table 3 summarizes the exciting new contributions of information quality research and citizen science to the broader information quality discipline.

The examples of research in citizen science provided above are not meant to be comprehensive and exhaustive. Instead, they illustrate the potential of this research domain for advancing knowledge and practice in IS related to IQ. As discussed earlier, citizen science information quality research actively investigates traditional aspects of information quality, novel concepts, methods for conducting research, and approaches to IS development and IQ management. Currently rooted in specific scientific disciplines, citizen science studies have much to offer to the traditional information quality work in IS. At the same time, as information quality is one of the core subject areas for the information systems research, citizen science promises unique opportunities for cross-pollination of theories and methods, and for exporting foundational knowledge from IS to sciences where practitioners actively seek collaborators but may be unaware of IS research community. In the next section, we further consider the role citizen science information quality research may play in connecting IS scholars with the broader scientific community.

## 6 Directions for Future Research

Citizen science is becoming a prolific area of research in sciences credited with an increasing number of scientific breakthroughs and discoveries, and indeed, its own journal - *Citizen Science: Theory and Practice* (Bonney et al. 2016). At the same time, citizen science presents a wicked theoretical and design challenge (Hevner et al. 2004) of ensuring that the data provided by ordinary people is of the highest quality possible for rigorous science. Extensive evidence has emerged to support the effectiveness of enlisting ordinary people in this endeavor, so these wicked challenges are part of what makes citizen science such a promising venue for information quality research. More and more interdisciplinary teams from sciences and humanities continue to devise innovative and ingenious methods for harnessing the power of the people. One can hardly find another example of the convergence of so many different disciplinary perspectives and ideas focused on ensuring high-quality data. This paper demonstrates the strong potential of citizen science to advance IQ and offer unique insights to the information systems community, identifying ways that IS researchers can contribute to this socially-important movement.

**Table 3** Sample contributions of IQ in citizen science research

| Information Quality Research Dimensions | IQ-enhancing methods employed in citizen science context |
|---|---|
| Training and evaluating learning and performance | • Require testing certification |
| | • Embedded assessments |
| | • Training data creators |
| | • Creating tasks which specifically do not require training |
| | • Attracting diverse crowds |
| Optimizing Tradeoffs | • Innovative approaches to data collection |
| | • Using effort information to standardize data |
| | • Using modeling approaches |
| | • Applying different analytic approaches |
| Human Factors in IQ | • Motivating participants |
| | • Mitigating known biases |
| Hardware factors | • Designing for multiple devices |
| | • Creating processes which are simple, flexible, easy to use |
| | • Allowing user to explore, experiment, tinker, and even "fail" |
| Social factors | • Supporting high user interactivity |
| | • Mitigating biases arising from high social interactivity |
| | • Detecting malfeasance and data sabotage |
| Challenging Traditional Assumptions | • Redefining information quality |
| | • Redefining the notion of domain expertise |
| | • Exploring novel solutions (e.g., gamification, AI, hybrid intelligence) |

Based on our review of research on citizen science IQ, several themes emerged:

- Citizen science is being actively pursued by scientists from different disciplines, who are rarely familiar with IS;
- Many of the disciplines engaged in citizen science view IQ issues from their own perspectives, using their own vocabularies, with limited incentive to contribute beyond their own areas of research (e.g., to IS literature); and
- Much of the research on citizen science is published in non-IS and non-IQ journals, making it likely to be overlooked by IQ scholars.

These observations suggest that citizen science can be an important area of focus for future research that increases contact and collaboration between researchers in IS and other scientific disciplines. Below, we identify several directions in information quality in citizen science to support the assimilation of citizen science research conducted in sciences into IS.

**Research Direction # 1** Survey of the state of the art in citizen science. As shown in this paper, there is a wealth of theoretical and practical knowledge that is being continuously generated in non-IS domains. Much of knowledge on IQ in citizen science comes from the broader scientific community, including biology, physics, geography, anthropology, heath care, and engineering. Their interdisciplinary approaches can provide inspiration for new IQ work in IS. We suggest IS researchers should continuously monitor citizen science literature (currently almost entirely outside IS) to find relevant results and solutions. We call on the due diligence of both researchers in IS and the reviewers of such work, as a literature review on citizen science is fundamentally incomplete if it includes IS publications alone.

To guide such surveys of the state of the art, researchers in IS can take advantage of the many reviews of citizen science and IQ published in domain journals in sciences and humanities. Indeed, such reviews have already commenced, but thus far they focus on the citizen science itself (e.g., Prestopnik et al. 2013; Wiggins et al. 2011) or the respective scientific disciplines in which citizen science is considered a research method (e.g., Lewandowski and Specht 2015). Publications such as Kosmala et al. (2016) focus on IQ in citizen science, but target domain research audiences (e.g., ecologists). More work is needed that surveys the ongoing research in the sciences from the perspective of the information quality tradition in IS, with the aim of assimilating these ideas and findings for the benefit of IS.

Among others, we encourage researchers to pursue the following questions:

- Which themes are emerging in IQ citizen science research in other disciplines? What are key trends and findings?
- Which IQ solutions are most actively pursued in other disciplines? Which of these are most promising for broader applications? Which require more analysis and investigations?
- What research areas are becoming saturated and what gaps are emerging? What conceptual areas have the most potential to result in significant insights for theory of IQ?
- Which findings hold the most practical relevance and impact? Which are most clearly translatable to other contexts?

**Research Direction # 2** Integration of research in citizen science with broader information quality research. Following from the previous research direction, once relevant studies have been identified and surveyed, their cumulative knowledge needs to be integrated with theories and methods in IS. This paper has attempted to begin demonstrating how citizen science research relates to existing knowledge in IS, including identification of what is novel and what is a new application of prior knowledge. But a more systematic effort is needed and this work is likely to be a challenging one. Given its inherent disciplinary roots, many studies in citizen science utilize language, concepts and pursue objectives very different from that of the IQ tradition in IS. Indeed, even the very term "citizen science" has multiple dimensions and interpretations (Eitzel et al. 2017; Kullenberg and Kasperowski 2016; Levy and Germonprez 2017; Wiggins and Crowston 2011), as indicated by a variety of labels each with its own accent and flavor, including volunteer surveys (Lewandowski and Specht 2015), citizen focus groups (Heikkila and Isett 2007), volunteered geographic information (Goodchild 2007), participatory Geoweb (Parfitt 2013), civic science (Clark and Illman 2001), and scientific crowdsourcing (Sabou et al. 2012). Yet, as argued earlier, this diversity makes this research area quite a productive source of original ideas.

To begin integrating citizen science with IS, we suggest researchers address the following questions and needs:

- What basic assumptions of IS IQ research does citizen science violate? Are these assumptions due for critical re-evaluation?
- Who uses the data generated by citizen science and how well does it achieve "fitness for purpose" for these data users? How does this compare to other instances of data production and use in IS studies?
- Find analogs of research problems studied in IS that also occur in citizen science. Once found, consider triangulating studies in IS context (e.g., on product reviews) with studies in citizen science context (e.g., descriptions of planets or galaxies based on digital imagery). This will not only strengthen the generalizability of specific studies but also move the field of IS and citizen science closer.
- Conducting systematic literature reviews of citizen science from IS perspective, including focused reviews on relevant sub-topics to improve the utility and depth of the analysis. This would include translation of disciplinary concepts into IS vocabulary via a nomological network.
- As citizen science is a rapidly moving field, such reviews should be repeated frequently. Here, a promising strategy is utilizing the novel machine-learning approaches for automatic literature review and knowledge integration currently under active development in IS (Endicott et al. 2017; Larsen and Bong 2016). Indeed, citizen science's vastness and rapid growth is easily "a big data of IQ research" and can motivate further development of automatic literature review and integration methods in IS.

**Research Direction # 3** Investigating and developing novel information quality approaches in citizen science. The IS community is increasingly engaging in design science research with the aim to produce innovative artifacts such as principles, models, methods, and instantiations (Goes 2014; Gregor and Hevner 2013; Rai 2017). With this paper we hope to increase awareness and recognition of citizen science by the broader IS community as a fertile area for design work. Citizen science presents many novel challenges exaggerated beyond their usual extent in traditional domains, creating opportunity for innovation. Indeed, several novel IQ propositions in IS originated in this context, including the impact of gamification on IQ (Prestopnik et al. 2013), the potentially detrimental role of domain expertise, the connection between conceptual modeling grammars and IQ dimensions (Lukyanenko et al. 2014b), and the motivational model that relates individual and social factors to accuracy and completeness (Nov et al. 2014). These and other advances are the tip of the proverbial iceberg for design work in this area.

We encourage researchers to adopt citizen science as their domain and investigate such pressing questions as:

- How can citizen science platforms be designed to allow maximally open citizen participation and to motivate contributions while delivering highest quality data to scientists?

- How can designs encourage freedom of expression, promoting novel perspectives and discoveries while ensuring that the data remains consistent and usable for science?
- How can designs mitigate participation and reporting biases (e.g., overrepresentation of charismatic species) and deliver more complete and representative data?
- How can citizen science be augmented with AI and hybrid intelligence to make it more effective?

**Research Direction # 4** Considering citizen science implications of research findings. Assimilation of citizen science in IS should not be unidirectional. IS researchers in IQ stand to enrich the citizen science literature by exporting IS knowledge. We call on researchers working on problems that are similar to citizen science to proactively consider implications of their studies for citizen science. While citizen science has unique characteristics, many of its features (e.g., anonymous users, diverse users, weak controls over information production) are present to various extents in other domains, including traditional corporate settings. This naturally includes areas conceptually adjacent to citizen science such as other types of crowdsourcing, social media, distributed virtual teams, and social networks where similar information quality challenges persist. For example, computer science and IS recently began to investigate crowdsourcing in small problem-solving tasks on platforms such as Amazon's Mechanical Turk (Amsterdamer et al. 2015; Deng et al. 2016; Franklin et al. 2011; Ipeirotis et al. 2010; Wang et al. 2012). While this environment differs from citizen science, with an explicit payment for work involving small tasks that are stripped of organizational context, the data production aspects of crowdsourcing share similarities with citizen science. So far, this line of work has been conducted in relative isolation (Lukyanenko and Parsons 2015a) and holds strong potential to advance citizen science research. In the process of developing citizen science projects, researchers routinely investigate, test, and implement novel approaches for effective engagement with ordinary people. Among other extensions, lessons from citizen science can be potentially applicable for corporate research and development initiatives which seek direct customer input. Companies are increasingly exploring these opportunities, such as Fortune 500 companies investing in digital platforms to monitor what potential customers are saying, understand customer reactions to products and services, use consumer feedback to design better products, and monitor market changes (Barwise and Meehan 2010; Brynjolfsson and McAfee 2014; Delort et al. 2011).

Indeed, some studies already demonstrate the potential of citizen science as a source of insight for data production in commercial settings. Ogunseye et al. (2017) triangulated the context of citizen science with product reviews on Amazon.com and showed that the same information quality patterns (related to decline in contribution diversity over time) hold in both environments. These findings support our contention that citizen science can be a fruitful source of insights for information quality issues in corporate and other settings.

Broadly, citizen science also presents an opportunity for IS to share foundational concepts, principles, and design knowledge that can support excellence and advancement in a broad range of scientific disciplines. This would have a variety of benefits for IS, including amplifying its relevance to practice.

As citizen science grapples with particularly daunting IQ challenges, it pioneers theoretical and practical approaches that can improve IQ in other domains. For example, in an analogy to citizen science, the role of telemedicine health IS is connecting patients (domain non-experts, familiar with own conditions) and medical professionals (domain experts) (Hailey et al. 2002). At the same time, since most of citizen science IQ research occurs in sciences, these researchers may be unfamiliar with the IS community, creating opportunity for IS research to bridge this gap by leveraging the knowledge of IQ to benefit other domains.

We thus encourage researchers to:

- Conduct cross-disciplinary literature reviews which identify common trends related to IQ (e.g., citizen science and participatory design, citizen science and open source software, citizen science and social media).
- Consider cross-disciplinary research questions such as: How does the effect of training differ in tasks that involve semantic (i.e., general, shared, as in typical citizen science) vs. episodic (i.e., personal, typical to social media) knowledge? How does the design of online training systems impact task performance?
- Leverage similarities in underlying technology to better understand the impact of designs on IQ. For example, addressing the question of how mobile environments (due to their limited screen space, but also novel affordances related to sensors and mobility) could impede or promote user engagement in citizen science vs social networking settings.

**Research Direction # 5** General framework on IQ that considers citizen science. Existing frameworks for understanding an improving information quality have focused on corporate data production (Blake and Shankaranarayanan 2015; Madnick et al. 2009; Wang et al. 1995a). To better integrate citizen science into the broader IQ tradition in IS, existing frameworks need to be extended to include UGC and citizen science. This is important as citizen science does not exhaust the IQ landscape of emerging technologies nor vice versa, so their inclusion should lead to more robust frameworks.

Indeed, some IQ topics have received relatively less attention in citizen science compared to other UGC contexts and other areas. For example, much of work that relates to the impact of content moderation and editing appears to originate in the context of online forums and wikis (Arazy et al. 2011; Delort et al. 2011; Kane and Ransbotham 2016; Liu and Ram 2011). Wikipedia is the prototypical case for work on collaborative content production and its impact on IQ (Arazy et al. 2011, 2016; Arazy and Kopak 2011; Liu and Ram 2011). Applying data mining to automatically parse through data to make UGC more usable by consumers tends to concentrate on social media and social networking data (Abbasi et al. 2016; Bifet and Frank 2010; Susarla et al. 2012). Other open collaboration contexts include online reviews (Pavlou and Dimoka 2006), Linked Open Data (Heath and Bizer 2011), micro-tasks and crowdsourcing markets (Deng et al. 2016; Ipeirotis et al. 2010), and blogs (Chau and Xu 2012; Wattal et al. 2010), all somewhat different from citizen science and promising their own unique insights. Thus, it is important to provide a comprehensive assessment of the entire space of UGC. Early work in this direction has been undertaken by Tilley et al. (2016), but considering the growth and diversity of UGC, more work of this kind is needed. The diversity of UGC landscape calls for the development of general framework for understanding IQ in UGC that would also position citizen science relative to other forms of information production.

Accordingly, we call on future studies to:

- Create a general IQ framework which would account for citizen science and other emerging technologies.
- Integrate the framework on emerging technologies with traditional IQ to provide a general and unified view of IQ management.

As follows from the illustration of some recent theoretical and practical contributions, along with many fertile areas for future work discussed above, citizen science is an area of exciting opportunities for intellectually stimulating and socially impactful research. As IQ is arguably one of the core problems of the information systems discipline, quality becomes a common thread that can connect IS with many other disciplines in sciences and humanities. As astronomers, biologists, physicists, anthropologists, and geographers constantly grapple with the daunting task of ensuring that the data provided by ordinary people meets their high standards, information systems research with its long IQ tradition, explicit IQ focus, and proven IQ solutions has the potential to become the foundation on which the citizen science architecture of the future is built.

In closing, although we focused on the opportunities in conducting information quality research on citizen science in IS, we reiterate the general call made by Levy and Germonprez (2017) for information systems researchers to adopt citizen science as a rich, promising, and socially important research context. For example, one traditional expertise area of information systems researchers is adoption of new technologies (Dwivedi et al. 2015; Larsen and Bong 2016; Sidorova et al. 2008; Venkatesh et al. 2007). Adoption issues are fraught with challenges in citizen science, as information quality concerns place real, or at times, psychological (Lewandowski and Specht 2015) barriers for scientists or policymakers interested in engaging with the public and using citizen-generated data in research or public policy decisions. Conducting research in citizen science also heeds the call within the IS discipline to conduct research that promotes or supports environmental sustainability through innovative information technologies (Goes 2014; Melville 2010; Seidel et al. 2013). We hope that this paper adds to this momentum and impels more work in IS to explore this new research frontier.

## References

Abbasi, A., Sarker, S., & Chiang, R. H. (2016). Big data research in information systems: Toward an inclusive research agenda. *Journal of the Association for Information Systems, 17*(2), 3.

Abbasi, A., Zhou, Y., Deng, S., & Zhang, P. (2018). Text analytics to support sense-making in social media: A language-action perspective. *MIS Quarterly, 42*(2), 1–38.

Allahbakhsh, M., Benatallah, B., Ignjatovic, A., Motahari-Nezhad, H. R., Bertino, E., & Dustdar, S. (2013). Quality control in crowdsourcing systems: Issues and directions. *IEEE Internet Computing, 17*(2), 76–81.

Alonso, O. (2015). Challenges with label quality for supervised learning. *Journal of Data and Information Quality, 6*(1), 2.

Amir B, Ralph P (2014) Proposing a theory of gamification effectiveness. Hyderabad: ACM.

Amsterdamer Y, Davidson SB, Kukliansky A, Milo T, Novgorodov S, Somech A (2015) Managing general and individual knowledge in crowd mining applications. Seventh Biennial Conference on Innovative Data Systems Research. Asilomar: CIDR.

Andriole, S. J. (2010). Business impact of web 2.0 technologies. *Communications of the ACM, 53*(12), 67–79.

Arazy, O., & Kopak, R. (2011). On the measurability of information quality. *Journal of the American Society for Information Science and Technology, 62*(1), 89–99.

Arazy, O., Nov, O., Patterson, R., & Yeo, L. (2011). Information quality in Wikipedia: The effects of group composition and task conflict. *Journal of Management Information Systems, 27*(4), 71–98.

Arazy, O., Daxenberger, J., Lifshitz-Assaf, H., Nov, O., & Gurevych, I. (2016). Turbulent stability of emergent roles: The dualistic nature of self-organizing knowledge coproduction. *Information Systems Research, 27*(4), 792–812.

Arazy, O., Kopak, R., & Hadar, I. (2017). Heuristic principles and differential judgments in the assessment of information quality. *Journal of the Association for Information Systems, 18*(5), 403.

Awal GK, Bharadwaj K (2017) Leveraging collective intelligence for behavioral prediction in signed social networks through evolutionary approach. *Information Systems Frontiers* 1–23. https://doi.org/10.1007/s10796-017-9760-4

Azzurro, E., Broglio, E., Maynou, F., & Bariche, M. (2013). Citizen science detects the undetected: The case of Abudefduf saxatilis from the Mediterranean Sea. *Management of Biological Invasions, 4*(2), 167–170.

Ballou, D. P., & Pazer, H. L. (1995). Designing information systems to optimize the accuracy-timeliness tradeoff. *Information Systems Research, 6*(1), 51.

Ballou, D. P., & Pazer, H. L. (2003). Modeling completeness versus consistency tradeoffs in information decision contexts. *IEEE Transactions on Knowledge and Data Engineering, 15*(1), 240–243.

Ballou D, Madnick S, Wang R (Winter 2003) Special section: Assuring information quality. Journal of Management Information Systems. 20(3):9–11.

Barwise, P., & Meehan, S. (2010). The one thing you must get right when building a brand. *Harvard Business Review, 88*(12), 80–84.

Batini C, Scannapieca M (2006) *Data quality: concepts, methodologies and techniques*. Berlin Heidelberg: Springer-Verlag

Batini, C., Rula, A., Scannapieco, M., & Viscusi, G. (2015). From data quality to big data quality. *Journal of Database Management, 26*(1), 60–82.

Bauer, M. W., Petkova, K., & Boyadjieva, P. (2000). Public knowledge of and attitudes to science: Alternative measures that may end the "science war". *Science, Technology & Human Values, 25*(1), 30–51.

Beath, C., Berente, N., Gallivan, M. J., & Lyytinen, K. (2013). Expanding the frontiers of information systems research: Introduction to the special issue. *Journal of the Association for Information Systems, 14*(4), 5.

Becker-Klein R, Peterman K, Stylinski C (2016) Embedded assessment as an essential method for understanding public engagement in citizen science. *Citizen Science: Theory and Practice, 1*(1), 1–10.

Belbin L (2011) The atlas of living Australia's spatial portal. 28–29.

Belden, O. S., Baker, S. C., & Baker, B. M. (2015). Citizens unite for computational immunology! *Trends in Immunology, 36*(7), 385–387.

Watson-Manheim, M. B., Chudoba, K. M., & Crowston, K. (2002). Discontinuities and continuities: A new way to understand virtual work. *Information Technology and People, 15*(3), 191–209.

Bifet, A., & Frank, E. (2010). Sentiment knowledge discovery in twitter streaming data discovery science. In B. Pfahringer, G. Holmes, & A. Hoffmann (Eds.), *Discovery science* (pp. 1–15). Heidelberg: Springer Berlin.

Bird, T. J., Bates, A. E., Lefcheck, J. S., Hill, N. A., Thomson, R. J., Edgar, G. J., Stuart-Smith, R. D., Wotherspoon, S., Krkosek, M., & Stuart-Smith, J. F. (2014). Statistical solutions for error and bias in global citizen science datasets. *Biological Conservation, 173*, 144–154.

Blake R, Shankaranarayanan G (2015) Data and information quality: Research themes and evolving patterns. *AMCIS 2015*, 1–10.

Bodker, S. (1996). Creating conditions for participation: Conflicts and resources in systems development. *Human Computer Interaction, 11*(3), 215–236.

Bonabeau, E. (2009). Decisions 2.0: The power of collective intelligence. *MIT Sloan Management Review, 50*(2), 45.

Bonney, R., Cooper, C. B., Dickinson, J., Kelling, S., Phillips, T., Rosenberg, K. V., & Shirk, J. (2009). Citizen science: A developing tool for expanding science knowledge and scientific literacy. *BioScience, 59*(11), 977–984.

Bonney, R., Shirk, J. L., Phillips, T. B., Wiggins, A., Ballard, H. L., Miller-Rushing, A. J., & Parrish, J. K. (2014a). Next steps for citizen science. *Science, 343*(6178), 1436–1437.

Bonney, R., Shirk Jennifer, L., Phillips, T. B., Wiggins, A., Ballard, H. L., Miller-Rushing, A. J., & Parrish, J. K. (2014b). Next steps for citizen science. *Science, 343*(6178), 1436–1437.

Bonney R, Cooper C, Ballard H (2016) The theory and practice of Citizen Science: Launching a new journal. *Citizen Science: Theory and Practice, 1*(1), 1–13.

Bowser A, Hansen D, Preece J, He Y, Boston C, Hammock J (2014) Gamifying citizen science: a study of two user groups. Proceedings of the companion publication of the 17th ACM conference on Computer supportedcooperative work & social computing (pp. 137–140). Baltimore: ACM.

Bowser A, Shilton K, Warrick E, Preece J (2017) Accounting for privacy in citizen science: Ethical research in a context of openness. Proceedings of the 2017 ACM Conference on Computer Supported Cooperative Work and Social Computing (pp. 2124–2136). Portland: ACM.

Boyajian, T., LaCourse, D., Rappaport, S., Fabrycky, D., Fischer, D., Gandolfi, D., Kennedy, G., Korhonen, H., Liu, M., & Moor, A. (2016). Planet hunters IX. KIC 8462852–where's the flux? *Monthly Notices of the Royal Astronomical Society, 457*(4), 3988–4004.

Brabham, D. C. (2013). *Crowdsourcing*. Cambridge: MIT Press.

Bratteteig, T., & Wagner, I. (2014). *Disentangling participation: Power and decision-making in participatory design*. New York: Springer International Publishing.

Brossard, D., Lewenstein, B., & Bonney, R. (2005). Scientific knowledge and attitude change: The impact of a citizen science project. *International Journal of Science Education, 27*(9), 1099–1121.

Brynjolfsson, E., & McAfee, A. (2014). *The second machine age: Work, progress, and prosperity in a time of brilliant technologies*. New York: WW Norton & Company.

Burgess H, DeBey L, Froehlich H, Schmidt N, Theobald E, Ettinger A, HilleRisLambers J, Tewksbury J, Parrish J (2017) The science of citizen science: Exploring barriers to use as a primary research tool. *Biological Conservation, 208*, 113–120

Burton-Jones, A., & Volkoff, O. (2017). How can we develop contextualized theories of effective use? A demonstration in the context of community-care electronic health records. *Information Systems Research, 28*(3), 468–489.

Calero, C., Caro, A., & Piattini, M. (2008). An applicable data quality model for web portal data consumers. *World Wide Web, 11*(4), 465–484.

Chau, M., & Xu, J. (2012). Business intelligence in blogs: Understanding consumer interactions and communities. *MIS Quarterly, 36*(4), 1189–1216.

Chittilappilly, A. I., Chen, L., & Amer-Yahia, S. (2016). A survey of general-purpose crowdsourcing techniques. *IEEE Transactions on Knowledge and Data Engineering, 28*(9), 2246–2266.

Choy, K., & Schlagwein, D. (2016). Crowdsourcing for a better world: On the relation between IT affordances and donor motivations in charitable crowdfunding. *Information Technology and People, 29*(1), 221–247.

Clark, F., & Illman, D. L. (2001). Dimensions of civic science introductory essay. *Science Communication, 23*(1), 5–27.

Clark, C., Wu, J., Pletsch, H., Guillemot, L., Allen, B., Aulbert, C., Beer, C., Bock, O., Cuéllar, A., & Eggenstein, H. (2017). The Einstein@home gamma-ray pulsar survey. I. Search methods, sensitivity, and discovery of new Young gamma-ray pulsars. *The Astrophysical Journal, 834*(2), 106.

Cohn, J. P. (2008). Citizen science: Can volunteers do real research? *AIBS Bulletin, 58*(3), 192–197.

Crall, A. W., Newman, G. J., Stohlgren, T. J., Holfelder, K. A., Graham, J., & Waller, D. M. (2011). Assessing citizen science data quality: An invasive species case study. *Conservation Letters, 4*(6), 433–442.

Crook T, Frasca B, Kohavi R, Longbotham R (2009) Seven pitfalls to avoid when running controlled experiments on the web. Proceedings of the 15th ACM SIGKDD international conference on Knowledge discovery and datamining (pp. 1105–1114). Paris: ACM

Daniel, F., Kucherbaev, P., Cappiello, C., Benatallah, B., & Allahbakhsh, M. (2018). Quality control in crowdsourcing: A survey of quality attributes, assessment techniques, and assurance actions. *ACM Computing Surveys CSUR, 51*(1), 7.

Daugherty, T., Eastin, M., & Bright, L. (2008). Exploring consumer motivations for creating user-generated content. *Journal of Interactive Advertising, 8*(2), 16–25.

Davenport, T., & Harris, J. (2017). *Competing on analytics: Updated, with a new introduction: The new science of winning*. Cambridge: Harvard Business Press.

Davis, A. K., & Howard, E. (2005). Spring recolonization rate of monarch butterflies in eastern North America: New estimates from citizen-science data. *Journal of the Lepidopterists' Society, 59*(1), 1–5.

Delaney, D. G., Sperling, C. D., Adams, C. S., & Leung, B. (2008). Marine invasive species: Validation of citizen science and implications for national monitoring networks. *Biological Invasions, 10*(1), 117–128.

DeLone, W. H., & McLean, E. R. (1992). Information systems success: The quest for the dependent variable. *Information Systems Research, 3*(1), 60–95.

Delort, J. Y., Arunasalam, B., & Paris, C. (2011). Automatic moderation of online discussion sites. *International Journal of Electronic Commerce, 15*(3), 9–30.

Deng, X., Joshi, K., & Galliers, R. D. (2016). The duality of empowerment and marginalization in microtask crowdsourcing: Giving voice to the less powerful through value sensitive design. *MIS Quarterly, 40*(2), 279–302.

Doan, A., Ramakrishnan, R., & Halevy, A. Y. (2011). Crowdsourcing systems on the world-wide web. *Communications of the ACM, 54*(4), 86–96.

Dwivedi, Y. K., Wastell, D., Laumer, S., Henriksen, H. Z., Myers, M. D., Bunker, D., Elbanna, A., Ravishankar, M., & Srivastava, S. C. (2015). Research on information systems failures and successes: Status update and future directions. *Information Systems Frontiers, 17*(1), 143–157.

Eitzel, M., Cappadonna, J. L., Santos-Lang, C., Duerr, R. E., Virapongse, A., West, S. E., Kyba, C. C. M., Bowser, A., Cooper, C. B., & Sforzi, A. (2017). Citizen science terminology matters: Exploring key terms. *Citizen Science: Theory and Practice, 2*(1).

Elevant K (2013) Why Share Weather? Motivational Model for" Share Weather" Online Communities and Three Empirical Studies. 46th Hawaii International Conference on System Sciences. https://doi.org/10.1109/HICSS.2013.26.

Endicott J, Larsen KR, Lukyanenko R, Bong CH (2017) Integrating Scientific Research: Theory and Design of Discovering Similar Constructs. *AIS SIGSAND Symposium* (Cincinnati, Ohio), 1–7.

ESCA (2015) 10 Principles of Citizen Science. *European Citizen Science Association. ECSA*. Retrieved (April 17, 2018), https://ecsa.citizen-science.net/engage-us/10-principles-citizen-science.

Eveleigh A, Jennett C, Lynn S, Cox AL (2013) I want to be a captain! i want to be a captain!: gamification in the old weather citizen science project. Proceedings of the First International Conference on Gameful Design, Research, and Applications (pp. 79–82). Toronto: ACM

Eveleigh A, Jennett C, Blandford A, Brohan P, Cox AL (2014) Designing for dabblers and deterring drop-outs in citizen science. Proceedings of the SIGCHI Conference on Human Factors in Computing Systems (pp. 2985–2994). Toronto: ACM.

Fielden, M. A., Chaulk, A. C., Bassett, K., Wiersma, Y. F., Erbland, M., Whitney, H., & Chapman, T. W. (2015). Aedes japonicus japonicus (Diptera: Culicidae) arrives at the most easterly point in North America. *Canadian Entomologist, 147*(06), 737–740.

Flanagin, A., & Metzger, M. (2008). The credibility of volunteered geographic information. *GeoJournal, 72*(3), 137–148.

Fortson L, Masters K, Nichol R, Borne K, Edmondson E, Lintott C, Raddick J, Schawinski K, Wallin J (2011) Galaxy Zoo: Morphological Classification and Citizen Science. *Advance Machine Learning Data Mineral Astronomy* 1–11.

Franklin MJ, Kossmann D, Kraska T, Ramesh S, Xin R (2011) CrowdDB: answering queries with crowdsourcing. *Proceedings 2011 ACM SIGMOD International Conference Management Data*. SIGMOD '11. (ACM, Athens, Greece), 61–72.

French, A. M., Guo, C., & Shim, J. P. (2014). Current status, issues, and future of bring your own device (BYOD). *CAIS, 35*, 10.

Galloway, A. W. E., Tudor, M. T., & Haegen, W. M. V. (2006). The reliability of citizen science: A Case study of Oregon white oak stand surveys. *Wildlife Society Bulletin, 34*(5), 1425–1429.

Garcia-Molina, H., Joglekar, M., Marcus, A., Parameswaran, A., & Verroios, V. (2016). Challenges in data crowdsourcing. *IEEE Transactions on Knowledge and Data Engineering, 28*(4), 901–911.

Garrity, E. J. (2001). Synthesizing user centered and designer centered IS development approaches using general systems theory. *Information Systems Frontiers, 3*(1), 107–121.

Gemino, A., & Wand, Y. (2004). A framework for empirical evaluation of conceptual modeling techniques. *Requirements Engineering, 9*(4), 248–260.

Ghezzi, A., Gabelloni, D., Martini, A., & Natalicchio, A. (2017). Crowdsourcing: A review and suggestions for future research. *International Journal of Management Reviews, 00*, 1–21.

Girres, J. F., & Touya, G. (2010). Quality assessment of the French OpenStreetMap dataset. *Transactions in GIS, 14*(4), 435–459.

Goes, P. B. (2014). Editor's comments: Design science research in top information systems journals. *MIS Quarterly, 38*(1), iii–viii.

Goodchild, M. (2007). Citizens as sensors: The world of volunteered geography. *GeoJournal, 69*(4), 211–221.

Goodchild, M. F., & Glennon, J. A. (2010). Crowdsourcing geographic information for disaster response: A research frontier. *International Journal of Digital Earth, 3*(3), 231–241.

Gray ML, Suri S, Ali SS, Kulkarni D (2016) The crowd is a collaborative network. Proceedings of the 19th ACM Conference on Computer-Supported Cooperative Work & Social Computing (pp. 134–147). San Francisco: ACM

Gregor, S., & Hevner, A. R. (2013). Positioning and presenting design science research for maximum impact. *MIS Quarterly, 37*(2), 337–355.

Gura, T. (2013). Citizen science: amateur experts. *Nature, 496*(7444), 259–261.

Hailey, D., Roine, R., & Ohinmaa, A. (2002). Systematic review of evidence for the benefits of telemedicine. *Journal of Telemedicine and Telecare, 8*(1), 1–7.

Haklay M (2013) Citizen science and volunteered geographic information: Overview and typology of participation. *Crowdsourcing Geographic Knowledge*. Dordrecht: Springer.

Haklay, M., & Weber, P. (2008). OpenStreetMap: User-generated street maps. *IEEE Pervasive Computing, 7*(4), 12–18.

Hand, E. (2010). People power. *Nature, 466*(7307), 685–687.

He, Y., & Wiggins, A. (2017). Implementing an environmental citizen science project: Strategies and concerns from educators' perspectives. *International Journal of Environment Science Education, 12*(6), 1459–1481.

Heath, T., & Bizer, C. (2011). *Linked data: Evolving the web into a global data space*. San Rafael: Morgan & Claypool Publishers.

Heikkila, T., & Isett, K. R. (2007). Citizen involvement and performance management in special-purpose governments. *Public Administration Review, 67*(2), 238–248.

Hevner, A., March, S., Park, J., & Ram, S. (2004). Design science in information systems research. *MIS Quarterly, 28*(1), 75–105.

Higgins, C. I., Williams, J., Leibovici, D. G., Simonis, I., Davis, M. J., Muldoon, C., van Genuchten, P., O'Hare, G., & Wiemann, S. (2016). Citizen OBservatory WEB (COBWEB): A generic infrastructure platform to facilitate the collection of citizen science data for environmental monitoring. *International Journal Spatial Data Infrastructures Research, 11*, 20–48.

Hochachka, W. M., Fink, D., Hutchinson, R. A., Sheldon, D., Wong, W. K., & Kelling, S. (2012). Data-intensive science applied to broad-scale citizen science. *Trends in Ecology & Evolution, 27*(2), 130–137.

Hopkins, N., Tate, M., Sylvester, A., & Johnstone, D. (2017). Motivations for 21st century school children to bring their own device to school. *Information Systems Frontiers, 19*(5), 1191–1203.

Howe, J. (2008). *Crowdsourcing: How the power of the crowd is driving the future of business*. New York: Random House.

Ipeirotis PG, Gabrilovich E (2014) Quizz: Targeted crowdsourcing with a billion (potential) users. Proceedings of the 23rd international conference on World wide web (pp. 143–154). Seoul: ACM.

Ipeirotis PG, Provost F, Wang J (2010) Quality management on amazon mechanical Turk. *Proceedings of the ACM SIGKDD Workshop Human Computer* (ACM), 64–67.

Irwin A (1995) *Citizen science: A study of people, expertise and sustainable development* (Psychology Press).

Irwin A, Michael M (2003) *Science, social theory & public knowledge* (McGraw-Hill Education (UK), London, UK).

Jabbari Sabegh MA, Lukyanenko R, Recker J, Samuel BM, Castellanos A (2017) Conceptual modeling research in information systems: What we now know and what we still do not know. *AIS SIGSAND*. (Cincinnati, Ohio), 1–12.

Jabbari MA, Lukyanenko R, Recker J, Samuel BM, Castellanos A (2018) Conceptual Modeling Research: Revisiting and Updating Wand and Weber's 2002 Research Agenda. *AIS SIGSAND*. (Syracuse, NY), 1–12.

Jackson, M., Weyl, O., Altermatt, F., Durance, I., Friberg, N., Dumbrell, A. J., Piggott, J., Tiegs, S., Tockner, K., & Krug, C. (2016). Chapter twelve-recommendations for the next generation of global freshwater biological monitoring tools. *Advances in Ecological Research, 55*, 615–636.

Janssens, A. C. J., & Kraft, P. (2012). Research conducted using data obtained through online communities: Ethical implications of methodological limitations. *PLoS Medicine, 9*(10), e1001328.

Jordan, J. M. (2017). Challenges to large-scale digital organization: The case of Uber. *Journal of Organization Design, 6*(1), 11.

Kane, G. C., & Ransbotham, S. (2016). Research note—Content and collaboration: An affiliation network approach to information quality in online peer production communities. *Information Systems Research, 27*(2), 424–439.

Kane, G. C., Alavi, M., Labianca, G. J., & Borgatti, S. (2014). What's different about social media networks? A framework and research agenda. *MIS Quarterly, 38*(1), 274–304.

Kapoor, K. K., Tamilmani, K., Rana, N. P., Patil, P., Dwivedi, Y. K., & Nerur, S. (2018). Advances in social media research: Past, present and future. *Information Systems Frontiers, 20*(3), 531–558.

Khatib, F., DiMaio, F., Cooper, S., Kazmierczyk, M., Gilski, M., Krzywda, S., Zabranska, H., Pichova, I., Thompson, J., & Popović, Z. (2011). Crystal structure of a monomeric retroviral protease solved by protein folding game players. *Nature Structural & Molecular Biology, 18*(10), 1175–1177.

Khoury, G. A., Liwo, A., Khatib, F., Zhou, H., Chopra, G., Bacardit, J., Bortot, L. O., Faccioli, R. A., Deng, X., & He, Y. (2014). WeFold: A coopetition for protein structure prediction. *Proteins Struct. Funct Bioinforma, 82*(9), 1850–1868.

Kim S, Mankoff J, Paulos E (2013) Sensors: evaluating a flexible framework for authoring mobile data-collection tools for citizen science. (ACM), 1453–1462.

Klein, H. K., & Hirschheim, R. (2001). Choosing between competing design ideals in information systems development. *Information Systems Frontiers, 3*(1), 75–90.

Kohavi R, Deng A, Frasca B, Walker T, Xu Y, Pohlmann N (2013) Online controlled experiments at large scale. (ACM, ACM SIGKDD international conference on Knowledge discovery and data mining), 1168–1176.

Korpela, E. J. (2012). SETI@ home, BOINC, and volunteer distributed computing. *Annual Review of Earth and Planetary Sciences, 40*, 69–87.

Kosmala, M., Wiggins, A., Swanson, A., & Simmons, B. (2016). Assessing data quality in citizen science. *Frontiers in Ecology and the Environment, 14*(10), 551–560.

Kullenberg, C., & Kasperowski, D. (2016). What is citizen science?–a scientometric meta-analysis. *PLoS One, 11*(1), e0147152.

Kyng, M. (1995). Making representations work. *Communications of the ACM, 38*(9), 46–55.

Larsen, K., & Bong, C. H. (2016). A tool for addressing construct identity in literature reviews and meta-analyses. *MIS Quarterly, 40*(3), 1–23.

Law E, Gajos KZ, Wiggins A, Gray ML, Williams AC (2017) Crowdsourcing as a Tool for Research: Implications of Uncertainty. Proceedings of the 2017 ACM Conference on Computer Supported Cooperative Work and Social Computing (pp. 1544–1561). Portland: ACM.

Lawrence, J. M. (2015). Rediscovery of the threatened Stoffberg widow butterfly, Dingana fraterna: The value of citizen scientists for African conservation. *Journal of Insect Conservation, 19*(4), 801–803.

Lee, Y. W. (2003). Crafting rules: Context-reflective data quality problem solving. *Journal of Management Information Systems, 20*(3), 93–119.

Lee, Y. W., Strong, D. M., Kahn, B. K., & Wang, R. Y. (2002). AIMQ: A methodology for information quality assessment. *Information Management, 40*(2), 133–146.

Levina, N., & Arriaga, M. (2014). Distinction and status production on user-generated content platforms: Using Bourdieu's theory of cultural production to understand social dynamics in online fields. *Information Systems Research, 25*(3), 468–488.

Levitin, A., & Redman, T. (1995). Quality dimensions of a conceptual view. *Information Processing and Management, 31*(1), 81–88.

Levy, M., & Germonprez, M. (2017). The potential for citizen science in information systems research. *Communications of the Association for Information Systems, 40*(1), 2.

Lewandowski, E., & Specht, H. (2015). Influence of volunteer and project characteristics on data quality of biological surveys. *Conservation Biology, 29*(3), 713–723.

Li G, Wang J, Zheng Y, Franklin M (2016) Crowdsourced data management: A survey. *IEEE Transactions on Knowledge and Data Engineering, 28*(9), 2296–2319.

Light A, Miskelly C (2014) Design for sharing. *North Umbria University Sustainable Social Network.*

Liu J, Ram S (2011) Who does what: Collaboration patterns in the wikipedia and their impact on data quality. *19th Workshop Information Technology System* 175–180.

Liu, K., Eatough, R., Wex, N., & Kramer, M. (2014). Pulsar–black hole binaries: Prospects for new gravity tests with future radio telescopes. *Monthly Notices of the Royal Astronomical Society, 445*(3), 3115–3132.

Losey, J., Perlman, J., & Hoebeke, E. (2007). Citizen scientist rediscovers rare nine-spotted lady beetle, Coccinella novemnotata, in eastern North America. *Journal of Insect Conservation, 11*(4), 415–417.

Louv, R., Dickinson, J. L., & Bonney, R. (2012). *Citizen science: Public participation in environmental research*. Ithaca: Cornell University Press.

Love J, Hirschheim R (2017) Crowdsourcing of information systems research. *European Journal of Information Systems* 1–18.

Lukyanenko R (2014) *An information modeling approach to improve quality of user-generated content*. PhD thesis. (Memorial University of Newfoundland, St. John's, NL Canada).

Lukyanenko R, Parsons J (2014) Using Field Experimentation to Understand Information Quality in User-generated Content. *CodeCon@MIT*.

Lukyanenko R, Parsons J (2015a) Beyond Task-Based Crowdsourcing Database Research. *AAAI Conf. Hum. Comput. Crowdsourcing AAAI HCOMP*. (San Diego, CA, USA), 1–2.

Lukyanenko, R., & Parsons, J. (2015b). Information quality research challenge: Adapting information quality principles to user-generated content. *ACM Journal of Data and Information Quality, 6*(1), 1–3.

Lukyanenko, R., & Parsons, J. (2018). Beyond micro-tasks: Research opportunities in observational crowdsourcing. *J Database Manag JDM, 29*(1), 1–22.

Lukyanenko R, Parsons J, Wiersma Y (2014a) The Impact of Conceptual Modeling on Dataset Completeness: A Field Experiment. *Proceedings of the International Conference Information System. ICIS*. 1–18.

Lukyanenko, R., Parsons, J., & Wiersma, Y. (2014b). The IQ of the crowd: Understanding and improving information quality in structured user-generated content. *Information Systems Research, 25*(4), 669–689.

Lukyanenko, R., Parsons, J., & Wiersma, Y. (2016a). Emerging problems of data quality in citizen science. *Conservation Biology, 30*(3), 447–449.

Lukyanenko, R., Parsons, J., Wiersma, Y., Sieber, R., & Maddah, M. (2016b). Participatory Design for User-generated Content: Understanding the challenges and moving forward. *Scandinavian Journal of Information Systems, 28*(1), 37–70.

Lukyanenko R, Parsons J, Wiersma Y, Maddah M (2019) Expecting the Unexpected: Effects of Data Collection Design Choices on the Quality of Crowdsourced User-generated Content. *Press MIS Quarterly*.

MacDonald, E. A., Donovan, E., Nishimura, Y., Case, N. A., Gillies, D. M., Gallardo-Lacourt, B., Archer, W. E., Spanswick, E. L., Bourassa, N., & Connors, M. (2018). New science in plain sight: Citizen scientists lead to the discovery of optical structure in the upper atmosphere. *Science Advances, 4*(3), eaaq0030.

Maddah M, Lukyanenko R, VanderMeer D (2018) Impact of Collection Structures and Type of Data on Quality of User-Generated Content. *WITS 2018*. (Santa Clara, CA), 1–10.

Madnick, S. E., Wang, R. Y., Lee, Y. W., & Zhu, H. (2009). Overview and framework for data and information quality research. *Journal of Data and Information Quality, 1*(1), 1–22.

Malone, T. W., Laubacher, R., & Dellarocas, C. (2010). Harnessing crowds: Mapping the genome of collective intelligence. *Sloan Management Review, 51*(3), 21–31.

McAfee, A., & Brynjolfsson, E. (2017). *Machine, platform, crowd: Harnessing our digital future*. New York: WW Norton & Company.

McKinley, D. C., Miller-Rushing, A. J., Ballard, H. L., Bonney, R., Brown, H., Cook-Patton, S. C., Evans, D. M., French, R. A., Parrish, J. K., & Phillips, T. B. (2016). Citizen science can improve conservation science, natural resource management, and environmental protection. *Biological Conservation*.

Melville, N. P. (2010). Information systems innovation for environmental sustainability. *MIS Quarterly, 34*(1), 1–21.

Miller, D. A., Weir, L. A., McClintock, B. T., Grant, E. H. C., Bailey, L. L., & Simons, T. R. (2012). Experimental investigation of false positive errors in auditory species occurrence surveys. *Ecological Applications, 22*(5), 1665–1674.

Miranda, S. M., Young, A., & Yetgin, E. (2016). Are social media emancipatory or hegemonic? Societal effects of mass media digitization in the Case of the SOPA discourse. *MIS Quarterly, 40*(3), 303–329.

Mylopoulos, J. (1998). Information modeling in the time of the revolution. *Information Systems, 23*(3–4), 127–155.

Nelson AG, Stylinski C, Becker-Klein R, Peterman K (2017) Exploring embedded assessment to document scientic inquiry skills within citizen science. *Citizen Inquiry.* (Routledge), 81–100.

Newman, C., Buesching, C. D., & Macdonald, D. W. (2003). Validating mammal monitoring methods and assessing the performance of volunteers in wildlife conservation—"Sed quis custodiet ipsos custodies?". *Biological Conservation, 113*(2), 189–197.

Newman, G., Zimmerman, D., Crall, A., Laituri, M., Graham, J., & Stapel, L. (2010). User-friendly web mapping: Lessons from a citizen science website. *International Journal of Geographical Information Science, 24*(12), 1851–1869.

Newman, G., Wiggins, A., Crall, A., Graham, E., Newman, S., & Crowston, K. (2012). The future of citizen science: Emerging technologies and shifting paradigms. *Frontiers in Ecology and the Environment, 10*(6), 298–304.

Nielsen, M. (2011). *Reinventing discovery: The new era of networked science*. Hoboken: Princeton University Press.

Nov O, Arazy O, Anderson D (2011) Dusting for science: motivation and participation of digital citizen science volunteers. *2011 IConference*: 68–74.

Nov, O., Arazy, O., & Anderson, D. (2014). Scientists@ home: What drives the quantity and quality of online citizen science participation. *PLoS One, 9*(4), 1–11.

Ogunseye S, Parsons J (2016) Can Expertise Impair the Quality of Crowdsourced Data? *SIGOPEN Development Workshop ICIS 2016*.

Ogunseye S, Parsons J, Lukyanenko R (2017) Do Crowds Go Stale? Exploring the Effects of Crowd Reuse on Data Diversity. *WITS 2017*. (Seoul, South Korea).

Osborn DA, Pearse JS, Roe CA, Magoon O, Converse H, Baird B, Jines B, Miller-Henson M (2005) Monitoring rocky intertidal shorelines: a role for the public in resource management. Conference Proceedings, California and the World Ocean 2002: Revisiting and Revising California's Ocean Agenda. Reston: American Society of Civil Engineers.

Palacios, M., Martinez-Corral, A., Nisar, A., & Grijalvo, M. (2016). Crowdsourcing and organizational forms: Emerging trends and research implications. *Journal of Business Research, 69*(5), 1834–1839.

Parfitt I (2013) *Citizen science in conservation biology: best practices in the geoweb era*. (Univ. of British Columbia).

Parsons, J., Lukyanenko, R., & Wiersma, Y. (2011). Easier citizen science is better. *Nature, 471*(7336), 37–37.

Pattengill-Semmens CV, Semmens BX (2003) Conservation and management applications of the reef volunteer fish monitoring program. *Coastal Monitoring Partnership* (Springer), 43–50.

Pavlou, P. A., & Dimoka, A. (2006). The nature and role of feedback text comments in online marketplaces: Implications for trust building, price premiums, and seller differentiation. *Information Systems Research, 17*(4), 392–414.

Petter, S., DeLone, W., & McLean, E. R. (2013). Information systems success: The quest for the independent variables. *Journal of Management Information Systems, 29*(4), 7–62.

Pipino, L. L., Lee, Y. W., & Wang, R. Y. (2002). Data quality assessment. *Communications of the ACM, 45*(4), 211–218.

Poblet M, García-Cuesta E, Casanovas P (2017) Crowdsourcing roles, methods and tools for data-intensive disaster management. *Information System Frontier* 1–17.

Preece J, Boston C, Maher ML, Grace K, Yeh T (2016) From Crowdsourcing Technology Design to Participatory Design and Back Again! 315.

Prestopnik N, Crowston K (2011) Gaming for (Citizen) Science: Exploring Motivation and Data Quality in the Context of Crowdsourced Science through the Design and Evaluation of a Social-Computational System. *IEEE International Conference E-Sci. Workshop EScienceW*. 1–28.

Prestopnik N, Crowston K (2012a) Citizen science system assemblages: understanding the technologies that support crowdsourced science. (ACM), 168–176.

Prestopnik N, Crowston K (2012b) Purposeful gaming & socio-computational systems: a citizen science design case. (ACM), 75–84.

Prestopnik, N. R., Crowston, K., & Wang, J. (2013). Data quality in purposeful games. *Computing, 1*, 2.

Rai, A. (2017). Diversity of design science contributions. *MIS Quarterly, 41*(1), iii–xviii.

Rana, N. P., Dwivedi, Y. K., Williams, M. D., & Weerakkody, V. (2015). Investigating success of an e-government initiative: Validation of an integrated IS success model. *Information Systems Frontiers, 17*(1), 127–142.

Redman, T. C. (1996). *Data quality for the information age*. Norwood: Artech House.

Renault H (2018) How the web is helping gamers find new spider species. *ABC News*. Retrieved (April 9, 2018), http://www.abc.net.au/news/2018-01-05/seven-new-spider-species-discovered-by-gamers/9303710.

Richards C, Thompson CW, Graham N (2014) Beyond designing for motivation: the importance of context in gamification. (ACM), 217–226.

Riesch H, Potter C (2013) Citizen science as seen by scientists: Methodological, epistemological and ethical dimensions. *Public Understanding Science* 0963662513497324.

Robinson MR, Imran A (2015) A Design Framework for Technology-Mediated Public Participatory System for the Environment.

Rotman D, Preece J, Hammock J, Procita K, Hansen D, Parr C, Lewis D, Jacobs D (2012) Dynamic changes in motivation in collaborative citizen-science projects. (ACM), 217–226.

Rowland K (2012) Citizen science goes "extreme." *Nature News*.

Sabou M, Bontcheva K, Scharl A (2012) Crowdsourcing research opportunities: lessons from natural language processing. (ACM), 17.

Scannapieco, M., Missier, P., & Batini, C. (2005). Data quality at a glance. *Datenbank-Spektrum, 14*, 6–14.

Schilthuizen M, Seip LA, Otani S, Suhaimi J, Njunjić I (2017) Three new minute leaf litter beetles discovered by citizen scientists in Maliau Basin, Malaysian Borneo (Coleoptera: Leiodidae, Chrysomelidae). *Biodiversity Data Journal* (5).

Schmeller, D. S., HENRY, P., Julliard, R., Gruber, B., Clobert, J., Dziock, F., Lengyel, S., Nowicki, P., Déri, E., & Budrys, E. (2009). Advantages of volunteer-based biodiversity monitoring in Europe. *Conservation Biology, 23*(2), 307–316.

Seidel, S., Recker, J. C., & Vom Brocke, J. (2013). Sensemaking and sustainable practicing: Functional affordances of information systems in green transformations. *Management Information Systems Quarterly, 37*(4), 1275–1299.

Shankaranarayanan, G., & Blake, R. (2017). From content to context: The evolution and growth of data quality research. *Journal of Data and Information Quality, 8*(2), 9.

Sheppard SA, Terveen L (2011) Quality is a verb: The operationalization of data quality in a citizen science community. (ACM), 29–38.

Sheppard S, Wiggins A, Terveen L (2014) Capturing quality: retaining provenance for curated volunteer monitoring data. 1234–1245.

Show, H. (2015). Rise of the citizen scientist. *Nature, 524*(7565), 265–265.

Siau, K., & Rossi, M. (2011). Evaluation techniques for systems analysis and design modelling methods – A review and comparative analysis. *Information Systems Journal, 21*(3), 249–268.

Sicari, S., Cappiello, C., De Pellegrini, F., Miorandi, D., & Coen-Porisini, A. (2016). A security-and quality-aware system architecture for internet of things. *Information Systems Frontiers, 18*(4), 665–677.

Sidorova A, Evangelopoulos N, Valacich JS, Ramakrishnan T (2008) Uncovering the intellectual core of the information systems discipline. *MIS Q.*:467–482.

Sieber, R. (2006). Public participation geographic information systems: A literature review and framework. *Annals of the Association of American Geographers, 96*(3), 491–507.

Silvertown, J. (2010). Taxonomy: Include social networking. *Nature, 467*(7317), 788–788.

Simpson E, Roberts S, Psorakis I, Smith A (2013) Dynamic bayesian combination of multiple imperfect classifiers. *Decision Making Imperfection*. (Springer), 1–35.

Simpson R, Page KR, De Roure D (2014) Zooniverse: observing the world's largest citizen science platform. *Proc. Companion Publ. 23rd International Conference World Wide Web Companion*. (International World Wide Web Conferences Steering Committee), 1049–1054.

Sorokin A, Forsyth D (2008) Utility data annotation with Amazon Mechanical Turk. *Comput. Vis. Pattern Recognit. Workshop 2008 CVPRW 08 IEEE Computer Social Conference On*. 1–8.

Sousa, L., de Mello, R., Cedrim, D., Garcia, A., Missier, P., Uchôa, A., Oliveira, A., & Romanovsky, A. (2018). VazaDengue: An information system for preventing and combating mosquito-borne diseases with social networks. *Information Systems, 75*, 26–42.

Stevens, M., Vitos, M., Altenbuchner, J., Conquest, G., Lewis, J., & Haklay, M. (2014). Taking participatory citizen science to extremes. *Pervasive Comput IEEE, 13*(2), 20–29.

Stewart, N., Ungemach, C., Harris, A. J., Bartels, D. M., Newell, B. R., Paolacci, G., & Chandler, J. (2015). The average laboratory samples a population of 7,300 Amazon mechanical Turk workers. *Judgment and Decision making, 10*(5), 479–491.

Strien, A. J., Swaay, C. A., & Termaat, T. (2013). Opportunistic citizen science data of animal species produce reliable estimates of distribution trends if analysed with occupancy models. *Journal of Applied Ecology, 50*(6), 1450–1458.

Strong, D. M., Lee, Y. W., & Wang, R. Y. (1997). Data quality in context. *Communications of the ACM, 40*(5), 103–110.

Susarla, A., Oh, J. H., & Tan, Y. (2012). Social networks and the diffusion of user-generated content: Evidence from YouTube. *Information Systems Research, 23*(1), 23–41.

Tan M, Tripathi N, Zuiker SJ, Soon SH (2010) Building an online collaborative platform to advance creativity. (IEEE).

Tang, J., Zhang, P., & Wu, P. F. (2015). Categorizing consumer behavioral responses and artifact design features: The case of online advertising. *Information Systems Frontiers, 17*(3), 513–532.

Theobald, E. J., Ettinger, A. K., Burgess, H. K., DeBey, L. B., Schmidt, N. R., Froehlich, H. E., Wagner, C., HilleRisLambers, J., Tewksbury, J., & Harsch, M. A. (2015). Global change and local solutions: Tapping the unrealized potential of citizen science for biodiversity research. *Biological Conservation, 181*, 236–244.

Tilly R, Posegga O, Fischbach K, Schoder D (2016) Towards a Conceptualization of Data and Information Quality in Social Information Systems. *Business & Information Systems Engineering, 59*(1) 3–21.

Tremblay, M. C., Dutta, K., & Vandermeer, D. (2010). Using data mining techniques to discover bias patterns in missing data. *Journal of Data and Information Quality, 2*(1), 2.

Venkatesh, V., Davis, F., & Morris, M. G. (2007). Dead or alive? The development, trajectory and future of technology adoption research. *Journal of the Association for Information Systems, 8*(4), 1.

Vitos M, Lewis J, Stevens M, Haklay M (2013) Making local knowledge matter: supporting non-literate people to monitor poaching in Congo. (ACM), 1–10.

Voosen P, 2018, Am 11:00 (2018) Update: NASA confirms amateur astronomer has discovered a lost satellite. *Science AAAS*. Retrieved (April 9, 2018), http://www.sciencemag.org/news/2018/01/amateur-astronomer-discovers-revived-nasa-satellite.

Wahyudi, A., Kuk, G., & Janssen, M. (2018). A process pattern model for tackling and improving big data quality. *Information Systems Frontiers, 20*(3), 457–469.

Wand, Y., & Weber, R. (1995). On the deep-structure of information-systems. *Information Systems Journal, 5*(3), 203–223.

Wand, Y., & Weber, R. (2002). Research commentary: Information systems and conceptual modeling - a research agenda. *Information Systems Research, 13*(4), 363–376.

Wang, R. Y. (1998). A product perspective on total data quality management. *Communications of the ACM, 41*(2), 58–65.

Wang, R. Y., & Strong, D. M. (1996). Beyond accuracy: What data quality means to data consumers. *Journal of Management Information Systems, 12*(4), 5–33.

Wang, R. Y., Reddy, M. P., & Kon, H. B. (1995a). Toward quality data: An attribute-based approach. *Decision Support Systems, 13*(3–4), 349–372.

Wang, R. Y., Storey, V. C., & Firth, C. P. (1995b). A framework for analysis of data quality research. *IEEE Transactions on Knowledge and Data Engineering, 7*(4), 623–640.

Wang J, Ghose A, Ipeirotis P (2012) Bonus, Disclosure, and Choice: What Motivates the Creation of High-Quality Paid Reviews? *International Conference Information System.*

Wattal, S., Schuff, D., Mandviwalla, M., & Williams, C. B. (2010). Web 2.0 and politics: The 2008 U.S. presidential election and an E-politics research agenda. *MIS Quarterly, 34*(4), 669–688.

Weeger A, Wang X, Gewald H, Raisinghani M, Sanchez O, Grant G, Pittayachawan S (2015) Determinants of intention to participate in corporate BYOD-Programs: The case of digital natives. *Information System Frontier* 1–17.

Wells, J. D., Valacich, J. S., & Hess, T. J. (2011). What signals are you sending? How website quality influences perceptions of product quality and purchase intentions. *MIS Quarterly, 35*(2), 373–396.

Wiersma, Y. F. (2010). Birding 2.0: Citizen science and effective monitoring in the web 2.0 world. *Avian Conservation Ecology, 5*(2), 13.

Wiggins A, Crowston K (2011) From Conservation to Crowdsourcing: A Typology of Citizen Science. *44th Hawaii International Conference System Science* 1–10.

Wiggins A, Crowston K (2014) Surveying the citizen science landscape. *First Monday 20*(1).

Wiggins A, He Y (2016) Community-based data validation practices in citizen science. (ACM), 1548–1559.

Wiggins A, Newman G, Stevenson RD, Crowston K (2011) Mechanisms for Data Quality and Validation in Citizen Science. *Computer Citizen Science Workshop*. (Stockholm, SE), 14–19.

Winter, S., Berente, N., Howison, J., & Butler, B. (2014). Beyond the organizational 'container': Conceptualizing 21st century sociotechnical work. *Information and Organization, 24*(4), 250–269.

Woo C (2011) The role of conceptual modeling in managing and changing the business. *International Conference Conceptual Model.* (Springer), 1–12.

Woolley, A. W., Chabris, C. F., Pentland, A., Hashmi, N., & Malone, T. W. (2010). Evidence for a collective intelligence factor in the performance of human groups. *Science, 330*(6004), 686–688.

Zhang, P., & Li, N. (2005). The intellectual development of human-computer interaction research: A critical assessment of the MIS literature (1990-2002). *Journal of the Association for Information Systems, 6*(11), 227–292.

Zhao, Y., & Zhu, Q. (2014). Evaluation on crowdsourcing research: Current status and future direction. *Information Systems Frontiers, 16*(3), 417–434.

Zhou, M. J., Lu, B., Fan, W. P., & Wang, G. A. (2018). Project description and crowdfunding success: An exploratory study. *Information Systems Frontiers, 20*(2), 259–274.

Zhu, H., & Wu, H. (2011). Quality of data standards: Framework and illustration using XBRL taxonomy and instances. *Electronic Markets, 21*(2), 129–139.

Zuboff, S. (1988). *In the age of the smart machine: The future of work and power*. New York: Basic Books.

Zwass, V. (2010). Co-creation: Toward a taxonomy and an integrated research perspective. *International Journal of Electronic Commerce, 15*(1), 11–48.

**Roman Lukyanenko** is an Assistant Professor in the Department of Information Technologies at HEC Montreal, Canada. Roman obtained his PhD from Memorial University of Newfoundland under the supervision of Dr. Jeffrey Parsons. Roman's research interests include conceptual modeling, information quality, citizen science, crowdsourcing, machine learning, design science research, classification theory, general ontology, and research methods (research validities, instantiation validity, and artifact sampling). Roman's work has been published in *Nature, MIS Quarterly, Information Systems Research, Journal of the Association for Information Systems, European Journal of Information Systems*, among others. Roman served as organizer, track chair and reviewer at numerous scientific conferences and as a Vice-President of the AIS Special Interest Group on Systems Analysis and Design.

**Andrea Wiggins** is an Assistant Professor of Information Systems and Quantitative Analysis in the College of Information Science and Technology at the University of Nebraska at Omaha. Her research focuses on sociotechnical systems for collaborative knowledge production and she serves as an advisor to citizen science initiatives across disciplines and around the world. Andrea's work has been published in *Science, BioScience, Ecology & Society, Human Computation, ACM Computing Surveys,* and the *Journal of the Association for Information Systems*, among others. She is a member of the editorial boards of *Frontiers in Ecology and the Environment* and *Citizen Science: Theory and Practice*.

**Holly Rosser** is a doctoral student in the College of Information Science and Technology at the University of Nebraska at Omaha. Her research interests include task instruction and tutorial design for online citizen science projects and platforms. She has published in the *Journal of Peer Production* and the proceedings of *HICSS 52* and *CSCW 2018*.